\documentclass[preprints,article,accept,pdftex,moreauthors]{Definitions/mdpi}

\firstpage{1} 
\pubvolume{1}
\issuenum{1}
\articlenumber{0}
\pubyear{2026}
\copyrightyear{2026}
\datereceived{ } 
\daterevised{ } 
\dateaccepted{ } 
\datepublished{ } 
\hreflink{https://doi.org/} 

\usepackage{braket}

\Title{Halo Nuclei from \textit{Ab Initio} Nuclear Theory}

\TitleCitation{Halo Nuclei from \textit{Ab Initio} Nuclear Theory}

\Author{Petr Navr\'atil $^{1,2,}$*\orcidA{}, Sofia Quaglioni $^{3}$, Guillaume Hupin $^{4}$, Michael Gennari $^{5}$ and Kostas Kravvaris $^{3}$}

\AuthorNames{Petr Navr\'atil, Sofia Quaglioni, Guillaume Hupin and Kostas Kravvaris}

\AuthorCitation{Navratil, P.; Quaglioni, S.; Hupin, G., Kravvaris, K.}

\address{%
$^{1}$ \quad TRIUMF, 4004 Wesbrook Mall, Vancouver, British Columbia V6T 2A3, Canada\\ 
$^{2}$ \quad University of Victoria, 3800 Finnerty Road, Victoria, British Columbia V8P 5C2, Canada\\
$^{3}$ \quad Lawrence Livermore National Laboratory, 7000 East Ave, Livermore, California 94550, USA\\
$^{4}$ \quad Universit\'e Paris-Saclay, CNRS/IN2P3, IJCLab, 91405 Orsay, France \\ 
$^{5}$ \quad Institut f\"{u}r Kernphysik, Johannes Gutenberg-Universit\"{a}t Mainz, 55128 Mainz, Germany 
}

\corres{Correspondence: navratil@triumf.ca}

\abstract{A realistic description of halo nuclei, characterized by low-lying breakup thresholds, requires a proper treatment of continuum effects. We have developed an \textit{ab initio} approach, the no-core shell model with continuum (NCSMC), capable of describing both bound and unbound states in light nuclei in a unified way. With chiral two- and three-nucleon interactions as the only input, we can predict structure and dynamics of halo and other light nuclei and, by comparing to available experimental data, test the quality of chiral nuclear forces. We review NCSMC calculations of weakly bound states and resonances of exotic halo nuclei $^6$He, $^8$B, $^{11}$Be, and $^{15}$C. For the latter, we discuss its production in the capture reaction $^{14}$C(n,$\gamma$)$^{15}$C. We highlight challenges of a description of $^6$He as a Borromean n-n-$^4$He system. 
Finally, we present calculations of excited states in $^{10}$Be exhibiting a one-neutron halo structure and a large scale no-core shell model investigation of $^{11}$Li as a precursor of a full n-n-$^9$Li NCSMC study.}


\begin{document}


\section{Introduction}

Halo nuclei are exotic weakly bound systems with extended single-nucleon or two-neutron density well beyond a tightly bound core. 
First discovered in $^{11}$Li nucleus from a series of interaction cross section measurements four decades ago~\cite{TA85}, where evidence suggests that this unstable nucleus exhibits a two-neutron halo structure around a central $^{9}$Li core and the distribution of halo neutrons extends to the size of a nucleus with mass number 200~\cite{TANIHATA2013215}. Similarly, the $^6$He nucleus is a prominent example of Borromean quantum halo, i.e., a weakly-bound state of three particles ($\alpha{+}n{+}n$) otherwise unbound in pairs, characterized by large probability of configurations within classically forbidden regions of space~\cite{Jensen2004}. Another prominent example is the $^{11}$Be featuring a parity-inverted ground state with extended neutron-$^{10}$Be $S$-wave halo~\cite{KELLEY201288}.

There has been a very significant experimental as well as theoretical effort to investigate halo nuclei~\cite{HansenJonson1987,Tanihata1996,TANIHATA2013215,Simon_2013,Riisager_2013,Hammer2017,Hagino2005,Redondo2008}. In this paper, we discuss \textit{ab initio}, or first-principles, description of halo nuclei. \textit{Ab initio} methods solve the many-body Schr\"{o}dinger equation for the system of $A$ nucleons interacting by forces derived within the chiral effective field theory (EFT) formalism~\cite{Weinberg1990}. These methods have been applied to halo nuclei in the past, e.g., the no-core shell model (NCSM)~\cite{Forssen2005,johnson2025challengesfirstprinciplesnuclearstructure} or nuclear lattice EFT~\cite{Shen2025,particles9010025}. As halo nuclei are characterized by low-lying breakup thresholds, a proper treatment of continuum effects is essential. Consequently, we focus on applications of the no-core shell model with continuum (NCSMC)~\cite{Baroni2013L,Baroni2013C,Navratil2016}, a method describing bound and unbound states in a unified way.  

The paper is organized as follows. In Section~\ref{sect:methods}, we describe the NCSMC method including the underlying NCSM~\cite{Barrett2013} approach. In Section~\ref{sect:results}, we review NCSMC applications to neutron and proton halo nuclei. In subsection~\ref{subsec:Be11}, we discuss the parity-inversion in $^{11}$Be and review published results for the $S$-wave halo of its ground state and present new results for the $P$-wave halo of its first excited state. In Subsection~\ref{subsec:C15}, we present new results for $^{15}$C which manifests a neutron $S$-wave halo ground state and discuss its production in the capture reaction $^{14}$C(n,$\gamma$)$^{15}$C relevant for astrophysics. Properties of $^8$B with a $P$-wave proton halo ground state are reviewed in Subsection~\ref{subsec:B8}. Our new results for excited neutron halo states in $^{10}$Be are presented in Subsection~\ref{subsec:Be10}. Applications of the three-body cluster NCSMC to the Borromean two-neutron halo nucleus $^6$He are reviewed in Subsection~\ref{subsec:He6}. Large-scale NCSM results for $^{11}$Li that also exhibits a Borromean two-neutron halo in its ground state are discussed in Subsection~\ref{subsec:Li11}. These calculations serve as a pre-requisite of a full three-body cluster NCSMC investigation. A concluding discussion is then given in Section~\ref{sect:discuss}.

\section{Materials and Methods}\label{sect:methods}

The starting point of our method is the microscopic Hamiltonian
\begin{equation}
\hat{H}=\frac{1}{A}\sum_{i<j=1}^A\frac{(\hat{\vec{p}}_i-\hat{\vec{p}}_j)^2}{2m} + \sum_{i<j=1}^A \hat{V}^{NN}_{ij} + \sum_{i<j<k=1}^A \hat{V}^{3N}_{ijk} \, ,\label{H}
\end{equation}
which describes nuclei as systems of $A$ non-relativistic point-like nucleons interacting through realistic inter-nucleon interactions. Modern theory of nuclear forces is based on the framework of chiral effective field theory (EFT)~\cite{Weinberg1990}, with the Lagrangian expanded in powers of $(Q/\Lambda_{\chi})^n$, where $Q$ is the external momentum and $\Lambda_{\chi}$ represents the hard scale of the theory of the order of 1 GeV. Such an expansion allows a systematic improvement of the interaction and provides a hierarchy of the nucleon-nucleon (NN) and many-nucleon interactions which naturally arise in a consistent scheme~\cite{OrRa94,EpNo02}.

As detailed in the next section, in the present work we apply several sets of chiral NN and chiral NN plus three-nucleon (3N) interactions consisting of NN interactions up to the third order (N$^2$LO)~\cite{Ekstrom2015}, fourth order (N$^3$LO)~\cite{Entem2003} or the fifth order (N$^4$LO)~\cite{Entem2017} in the chiral expansion and a 3N interaction up to the N$^2$LO order, in some cases including a subleading contact interaction~\cite{Girlanda2011}, regulated by local regulators~\cite{Navratil2007}, non-local regulators~\cite{Ekstrom2015} or combination of both~\cite{Soma2000}. The interaction parameters, the low-energy constants (LECs), are determined typically in $A{=}2,3,4$ nucleon systems, although properties of medium mass nuclei can also be considered ~\cite{Ekstrom2015}. 

A faster convergence of our NCSMC calculations is obtained by softening the chiral interaction through the similarity renormalization group (SRG) technique~\cite{Wegner1994,Bogner2007,Jurgenson2009}. The SRG unitary transformation induces many-body forces that we include up to three-body level. The four- and higher-body induced terms are small at the $\lambda_{\mathrm{SRG}}{\approx}1.8-2.0$ fm$^{-1}$ range of the resolution scale used in present calculations.

In the NCSMC~\cite{Baroni2013L,Baroni2013C,Navratil2016}, the many-body scattering problem is solved by expanding the wave function on continuous microscopic-cluster states, describing the relative motion between target and projectile nuclei and discrete square-integrable states, describing the static composite nuclear system. The idea behind this generalized expansion is to augment the microscopic cluster model, which enables the correct treatment of the wave function in the asymptotic region, with short-range many-body correlations that are present at small separations, mimicking various deformation effects that might take place during the reaction process. The NCSMC wave function for the $A$-nucleon system is represented as
\begin{align}
\ket{\Psi^{J^\pi}_{A, M_T}} = &  \sum_\lambda c^{J^\pi}_\lambda \ket{A \, \lambda J^\pi T M_T} 
+\sum_{\nu}\!\! \int \!\! dr \, r^2 
                 \frac{\gamma^{J^\pi}_{\nu}(r)}{r}
                 {\mathcal{A}}_\nu \ket{\Phi^{J^\pi}_{\nu r, M_T}} \,.\label{ncsmc_wf}
\end{align}
The first term of Eq. (\ref{ncsmc_wf}) consists of an expansion over square-integrable eigenstates of the composite nucleus indexed by $\lambda$. The second term corresponding to an expansion over the antisymmetrized channel states in the spirit of the resonating group
method (RGM)~\cite{wildermuth1977unified,TANG1978167,Quaglioni2009} is given by
\begin{align}
\ket{\Phi^{J^\pi}_{\nu r, M_T}} = &\Big[ \big( \ket{A-a \, \lambda_1 J_1^{\pi_1}T_1 M_{T_1}} \ket{a \, \lambda_2 J_2^{\pi_2}T_2 M_{T_2}} \big)^{(s)}
Y_\ell(\hat{r}_{A-a,a}) \Big]^{(J^{\pi})}_{M_T} 
\, \;\frac{\delta(r{-}r_{A-a,a})}{rr_{A-a,a}} \, .
\label{eq_rgm_state}
\end{align}
The $\nu$ index represents all the quantum numbers on the right-hand side not appearing on the left-hand side and the subscript $M_T{=}M_{T_1}{+}M_{T_2}$ is the isospin projection, i.e., $(Z-N)/2$. The coordinate $\vec{r}_{A-a,a}$ in Eq.(\ref{eq_rgm_state}) is the separation distance between the $(A{-}a)$-nucleon target and the $a$-nucleon projectile. It should be noted that the sum in the second term of Eq.~(\ref{ncsmc_wf}) comprises in general all the mass partitions involved in the formation of the composite system including three- or higher-body clusters, see the discussion below and in Subsection~\ref{subsec:He6}. For technical reasons, the NCSMC calculations are typically limited to one or two (e.g., energetically lowest~\cite{Hupin2019} or charge-exchange~\cite{Gysbers2024}) mass partitions.

The translationally invariant eigenstates of the aggregate ($A$), target ($A{-}a$), and projectile ($a$) nuclei are all obtained by means of the NCSM~\cite{Navratil2000L,Navratil2000,Barrett2013} using a basis of many-body harmonic oscillator (HO) wave functions with the same frequency, $\Omega$, and maximum number of particle excitations $N_{\rm max}$ from the lowest Pauli-allowed many-body configuration. The case of $a{=}1$ is trivial, the projectile wave function is simply $\ket{\frac{1}{2}^+ \frac{1}{2}\, \frac{1}{2}}$ ($\ket{\frac{1}{2}^+ \frac{1}{2}\, {-}\frac{1}{2}}$) for proton (neutron). 

We note that the approximate isospin quantum number $T$ is included in the NCSM eigenstates in Eqs.~(\ref{ncsmc_wf}) and (\ref{eq_rgm_state}) as they provide a useful information. However, in general we do not couple the target and projectile isospins to the total isospin. An exception to this is the $^{11}$B calculation discussed in Subsection~\ref{subsec:Be11}.

The discrete expansion coefficients  $c_{\lambda}^{J^{\pi}}$ and the continuous relative-motion amplitudes $\gamma_{\nu}^{J^{\pi}}(r)$ are the solution of the generalized eigenvalue problem derived by representing the Schr\"{o}dinger equation in the model space of the expansions (\ref{ncsmc_wf})~\cite{Navratil2016}. The resulting NCSMC equations are solved by the coupled-channel R-matrix method on a Lagrange mesh~\cite{Descouvemont2010,Hesse1998}.

An intuitive interpretation of wave functions of halo nuclei is provided by the overlap of the full solution for the eigenstate $\ket{\Psi^{J^\pi}_{A, M_T}}$ in Eq. (\ref{ncsmc_wf}) with the cluster portion $\ket{\Phi^{J^\pi}_{\nu r, M_T}}$ given by%
\begin{equation}
r\bra{\Phi^{J^\pi}_{\nu r, M_T}}\mathcal{A}_\nu\ket{\Psi^{J^\pi}_{A, M_T}} \,
\label{eq:NCSMC_formf}
\end{equation}
called cluster form factor. Integrating the cluster form factor squared, one obtains the spectroscopic factor. It is straightforward to evaluate cluster form factors within the NCSMC~\cite{Calci2016} as well as within the NCSM (i.e., with $\ket{\Psi^{J^\pi}_{A, M_T}}$ replaced by $\ket{A \, \lambda J^\pi T M_T}$ in Eq.~(\ref{eq:NCSMC_formf}))~\cite{Navratil2004cl}.

Additonal important characteristics of nuclear bound states, and the halos states in particular, are the asymptotic normalization coefficients (ANCs). In binary cluster scattering, these coefficients parameterize the bound state asymptotics of the nuclear wave function, i.e., the Coulombic tail, and are accessible quantities in scattering experiments. In a given partial wave channel, these coefficients are defined as
\begin{align}
    \lim_{ \vert \vec{r} \vert \rightarrow \infty } \Psi_{lm} ( \vec{r} \, )
    = C_{l} \ \frac{ W_{ -\eta_B, \; l+1/2} ( 2 \kappa \vert \vec{r} \vert ) }{ \vert \vec{r} \vert } \ Y_{lm} ( \hat{r} )
    \qquad \qquad
    \eta_{B} = \frac{ Z_{a} Z_{b} e^{2} \mu }{ \hbar^{2} \kappa } 
    \quad , \label{eq:ANC}
\end{align}
where $ C_{l} $ is the ANC, $ W $ is the Whittaker function, $ \eta_{B} $ is the Coulomb-Sommerfeld parameter, $ \mu $ is the reduced mass of the two-component system, and $ E = - \hbar^{2} \kappa^{2} / 2 \mu $ is the bound state energy of the system with respect to the breakup threshold.

The NCSMC can also be extended to describe systems dominated by three-body (and in principle, several-body) breakup channels by coupling discrete NCSM eigenstates with microscopic three-cluster continuum states, enabling a unified treatment of short-range many-body correlations and correct three-body asymptotics. To accomplish this, it is convenient to introduce an appropriate set of Jacobi relative coordinates among the clusters. For a system of three clusters with mass numbers $ A - a_{23} $, $ a_2 $, and $ a_3 $ (with $ a_{23} = a_2 + a_3 < A $), one possible choice is  
\begin{equation}
        \vec\eta_{1,23}  = (\eta_{1,23}, \theta_{\eta_{1,23}}, \phi_{\eta_{1,23}})  \label{eq:etay}
         = \sqrt{\frac{a_{23}}{A(A-a_{23})}}  \sum_{i=1}^{A-a_{23}} \vec{r}_i 
         - \sqrt{\frac{A-a_{23}}{A\,a_{23}}} \sum_{j=A-a_{23}+1}^A \vec{r}_j \,,
\end{equation}
which is the relative vector proportional to the displacement between the center of mass of the first cluster and that of the residual two fragments, and  
\begin{equation}
        \vec\eta_{23}  = (\eta_{23}, \theta_{\eta_{23}}, \phi_{\eta_{23}}) \label{eq:etax} 
         = \sqrt{\frac{a_3}{a_{23}\,a_2}}  \sum_{i=A-a_{23}+1}^{A-a_3} \vec{r}_i 
         - \sqrt{\frac{a_2}{a_{23}\,a_3}} \sum_{j=A-a_3+1}^A \vec{r}_j\,,      
\end{equation}
which is the relative coordinate proportional to the distance between the centers of mass of clusters 2 and 3. Here, $ \vec{r}_i $ denotes the position vector of the $ i $th nucleon.

\begin{figure}[H]
\includegraphics[width=6cm]{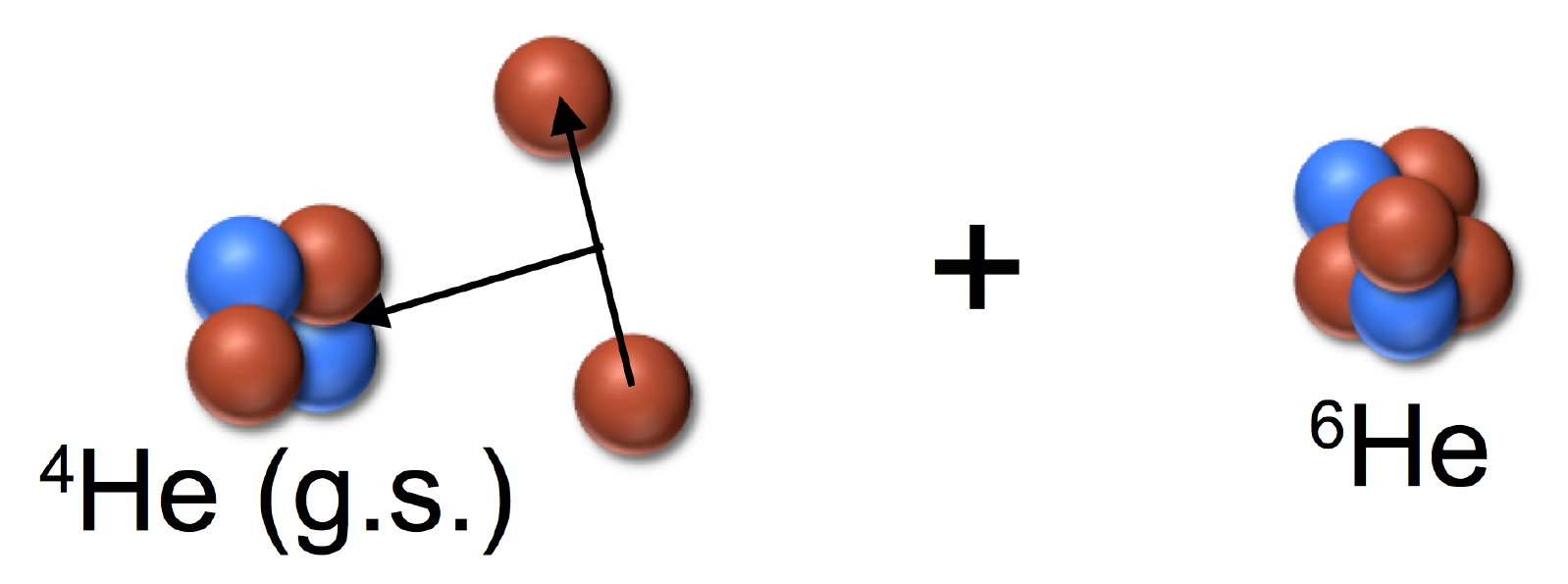}
\caption{Schematic depiction of the NCSMC basis expansion for $^6$He showing the NCSM $^6$He part and the three-body cluster part consisting of the $^4$He ground state and two neutrons.}
\label{fig:He6_3b-cluster}
\end{figure}

Within such three-cluster coordinate system, the ansatz for the many-body wave function of a Borromean halo nucleus such as, e.g., $^6$He can be written analogously to Eq.~\eqref{ncsmc_wf} (see Fig.~\ref{fig:He6_3b-cluster}),  
\begin{align}
\label{eq:trialwf}
	|\Psi^{J^\pi}_{A, M_T}\rangle 
	&= \sum_{\lambda} c^{J^\pi}_{\lambda}\,|A\lambda J^{\pi} M_T\rangle 
	+ \sum_{\nu K} \int d\rho\, \rho^5 \frac{\gamma_{K\nu}^{J^\pi}(\rho)}{\rho^{5/2}}\, \hat{\mathcal A}_\nu\, |\Phi^{J^\pi}_{\nu K \rho, M_T} \rangle\,,
\end{align}
where the RGM channel states for three clusters are given by  
\begin{align}
	|\Phi^{J^\pi}_{\nu K \rho, M_T} \rangle 
	= &\Big[\Big(|A-a_{23}\, \alpha_1 I_1^{\pi_1} M^T_1\rangle 
	\left(|a_2\, \alpha_2 I_2^{\pi_2} M^T_2\rangle\, |a_3\, \alpha_3 I_3^{\pi_3} M^T_3\rangle \right)^{(s_{23)}}_{M^T_{23}}\Big)^{(S)}_{M_T} \\ \nonumber
	&\times\; {\mathcal Y}_{L}^{K\ell_x\ell_y}(\Omega_\eta)\Big]^{(J^{\pi})}_{M_T} 
	 \frac{\delta(\rho - \rho_\eta)}{\rho^{5/2}\;\rho_\eta^{5/2}}\,.
	\label{eq:3bchannelHH}	
\end{align}

Here, we have introduced the hyperradial and hyperangular coordinates  
$ \rho_\eta = \sqrt{\eta_{23}^2 + \eta_{1,23}^2} $ and  
$ \alpha_\eta = \arctan\left(\frac{\eta_{23}}{\eta_{1,23}} \right) $,  
and the set of hyperangles  
$ \Omega_\eta = (\alpha_\eta, \theta_{\eta_{1,23}}, \phi_{\eta_{1,23}}, \theta_{\eta_{23}}, \phi_{\eta_{23}}) $.  
The functions $ {\mathcal Y}_{L}^{K\ell_x\ell_y}(\Omega_\eta) $ are hyperspherical harmonics with total orbital angular momentum $ L $ and hyperangular momentum $ K $, and $ \gamma_{K\nu}^{J^\pi}(\rho) $ are unknown hyperradial amplitudes. The remaining notation follows that introduced in Eq.~\eqref{ncsmc_wf}. For a detailed discussion of the three-cluster RGM formalism, we refer the interested reader to, e.g., Refs.~\cite{Quaglioni2013,Quaglioni2018}.

\section{Results}\label{sect:results}

\subsection{Parity inversion in $^{11}$Be}\label{subsec:Be11}

The theoretical understanding of exotic neutron-rich nuclei constitutes a tremendous challenge. These systems often cannot be explained by mean-field approaches and contradict the regular shell structure. The spectrum of $^{11}$Be has some very peculiar features. The $1/2^+$ ground state (g.s.) is loosely bound by 502 keV with respect to the n+$^{10}$Be threshold and is separated by only 320 keV from its parity-inverted $1/2^-$ partner, which would be the expected g.s. in the standard shell-model picture. Both these states exhibit a distinct $^{10}$Be+n halo structure. An accurate description of this complex spectrum is anticipated to be sensitive to the details of the nuclear force, such that a precise knowledge of the NN and also the 3N interaction, desirably obtained from first principles, is crucial. At the same time, an explicit treatment of continuum effects is indispensable. 

\begin{figure}[H]
\includegraphics[width=0.9\textwidth]{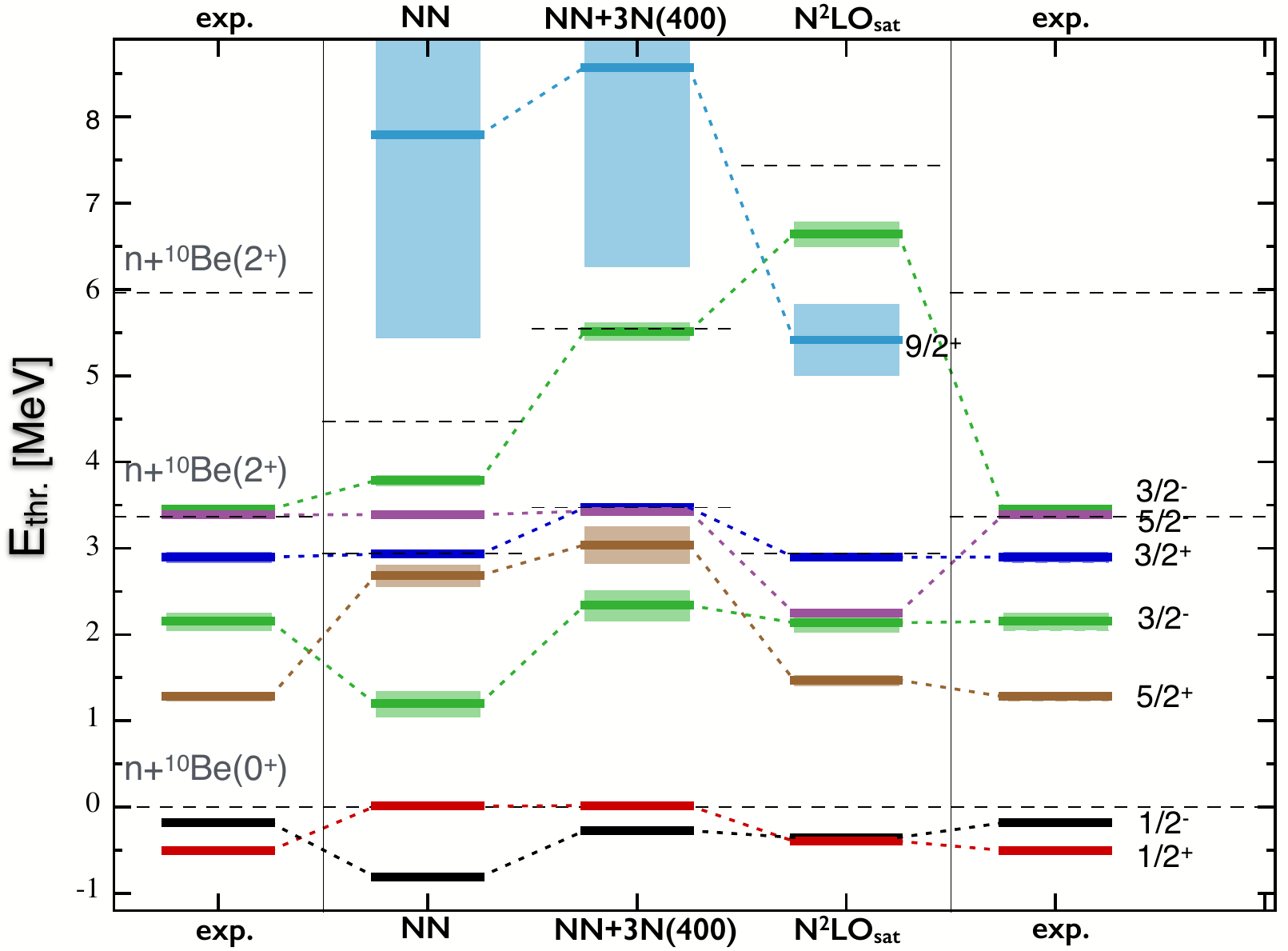}
\caption{NCSMC spectrum of $^{11}$Be with respect to the $n+^{10}$Be threshold for different chiral interactions compared to experiment. Dashed black lines indicate the energies of the $^{10}$Be states. Light boxes indicate resonance widths.}
\label{fig:Be11_spectrum}
\end{figure}
 The first NCSMC investigation of $^{11}$Be has been reported in Ref.~\cite{Calci2016} using several sets of chiral NN and 3N interactions. The calculations included the lowest three states of $^{10}$Be ($0^+_{1}, 2^+_1, 2^+_2$) and a range of NCSM eiegenstates of $^{11}$Be of both parities. A significant sensitivity to the chiral nuclear forces have been found. In Fig.~\ref{fig:Be11_spectrum}, we compare calculated levels to experiment, all relative to the n+$^{10}$Be($0^+$) threshold, for three sets of interactions. First, only the chiral N$^3$LO NN interaction~\cite{Entem2003} is used with no chiral 3N force (SRG induced 3N contributions are included in all calculations). The $1/2^-$ ground state is predicted incorrectly with the $1/2^+$ state at the threshold and the $3/2^-_1$ and $5/2^+$ states inverted as well. Adding the chiral 3N force with a local regulator with the cutoff of 400 MeV~\cite{Roth2012} improves the spectrum although the incorrect level orderings remain. On the contrary, the spectrum with the NN + 3N interaction simultaneously fitted to few-nucleon systems and medium mass nuclei, named N$^2$LO$_{\rm SAT}$~\cite{Ekstrom2015}, successfully achieves the parity inversions between the $3/2^-_1$ and $5/2^+$ resonances and, albeit marginally, for the bound states. The low-lying spectrum is significantly improved and agrees well with the experiment, presumably due to the more accurate description
of long-range properties caused by the fit of the interaction to radii of $p$-shell nuclei. On the other hand, the strongly overestimated splitting between the $3/2^-_2$ and $5/2^-$ states hints at deficiencies of this interaction, which might originate from a too large splitting of the $p_{1/2}{-}p_{3/2}$ subshells.  

\begin{figure}[H]
\centering
\includegraphics[width=\textwidth]{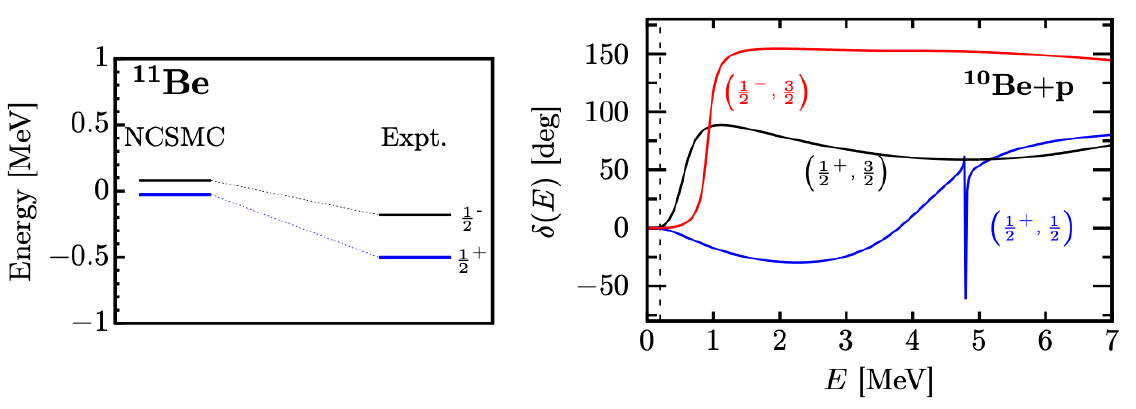}
\caption{(a) NCSMC calculated and experimental levels of $^{11}$Be. Only states corresponding to experimentally bound states with respect to the
$^{10}$Be$+n$ threshold (zero energy in the figure) are shown. (b) NCSMC-calculated $^{10}$Be$+p$ eigenphase shifts. The vertical dashed line indicates the experimentally predicted location of the ($1/2^+$, 1/2) resonance at 197 keV. The NN-N$^4$LO+3N$_{\rm lnl}$ interaction was used. Adapted from Ref.~\cite{Atkinson2022}.}
\label{fig:Be11_pBe10_N4LO}
\end{figure}
More recently, $^{11}$Be has been investigated with the NCSMC in the context of the $\beta$-delayed proton emission~\cite{Atkinson2022}. In that work, the chiral N$^4$LO NN interaction~\cite{Entem2017} combined with an N$^2$LO 3N interaction with simultaneous local and nonlocal regularization was used. Originally introduced in Ref.~\cite{Gysbers2019}, it is denoted as NN-N$^4$LO + 3N$_{\rm lnl}$. The SRG evolution was applied with 3N induced terms included. As shown in the left panel of Fig.~\ref{fig:Be11_pBe10_N4LO}, with this interaction the parity inversion in the ground state of $^{11}$Be has been comfortably reproduced. Moreover, the isospin analog states in $^{11}$B have been investigated within NCSMC considering the $^{10}$Be+p cluster. As seen in the right panel of Fig.~\ref{fig:Be11_pBe10_N4LO}, the parity inversion is also reproduced in the $T{=}3/2$ resonances in agreement with experiment.

\begin{figure}[hbt!]
\centering
\subfloat[\centering]{\includegraphics[width=8.4cm]{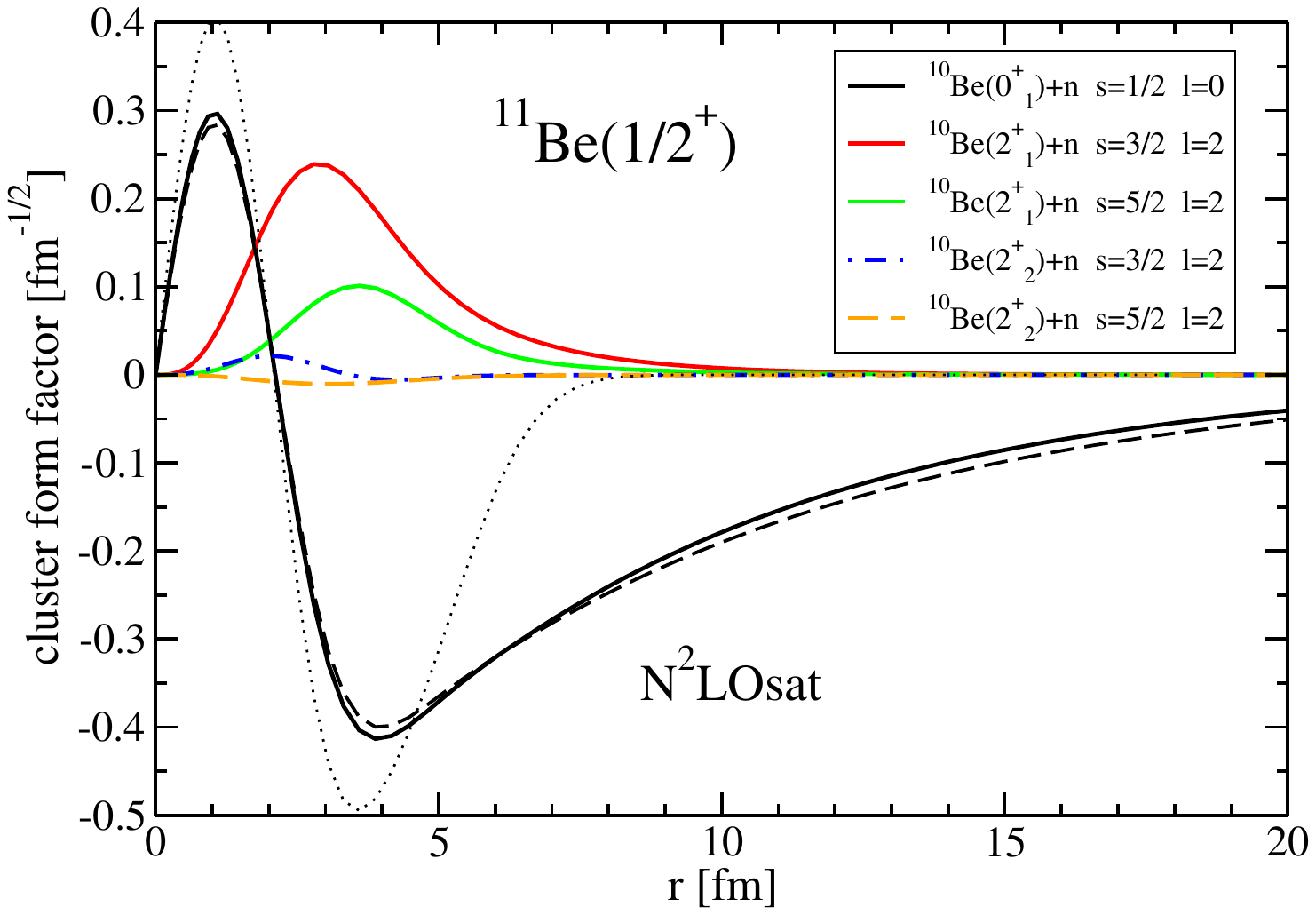}}
\subfloat[\centering]{\includegraphics[width=8.4cm]{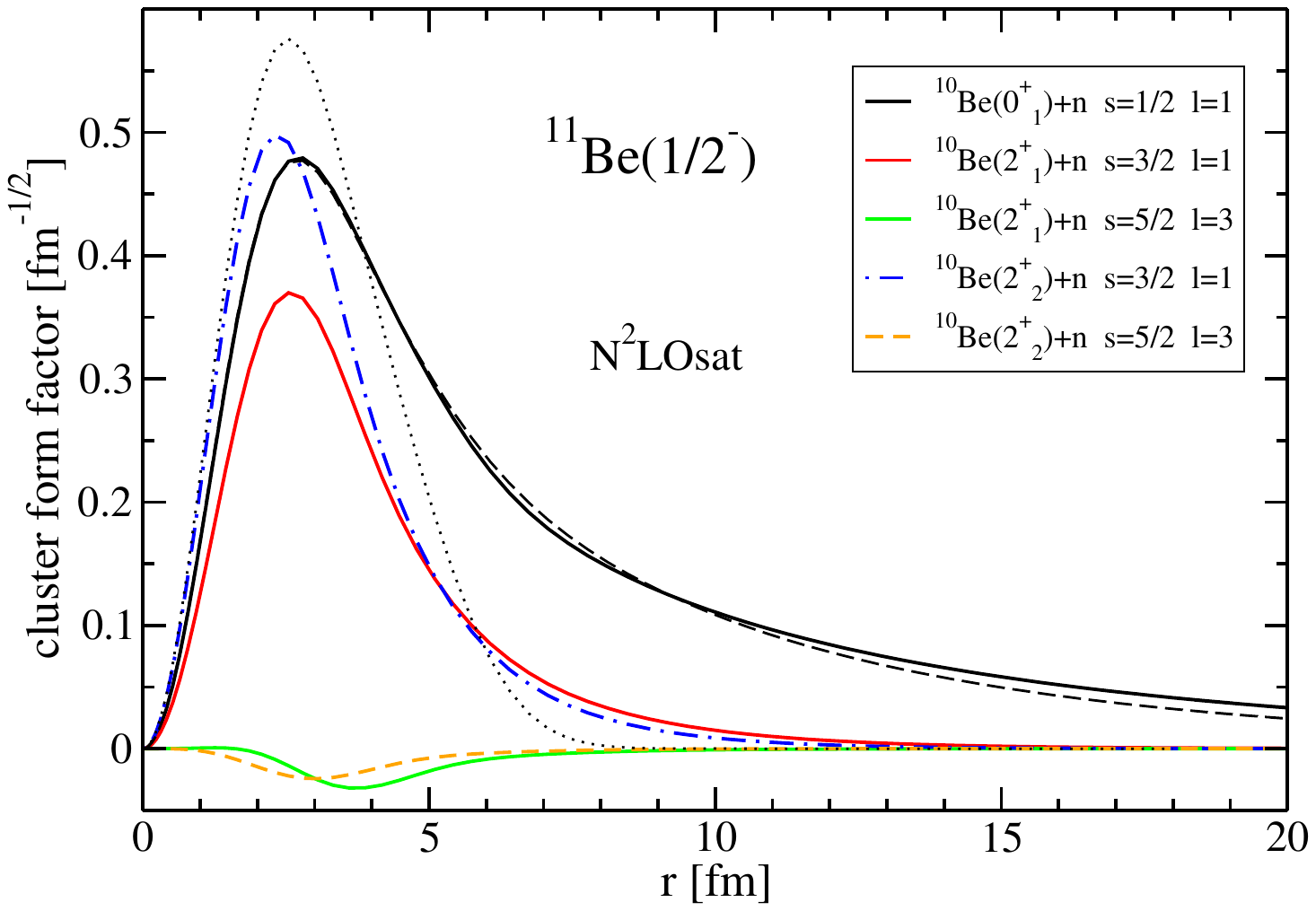}}
\caption{Cluster form factors of $^{11}$Be $1/2^+$ (a) and $1/2^-$ (b) states obtained with the N$^2$LOsat interaction. The solid lines show the NCSMC-pheno results, the black dashed (dotted) lines are the NCSMC (NCSM) $S$-wave (a) and $P$-wave (b) results. The legend columns show the $^{10}$Be eigenstate, the channel spin $s$ and the relative orbital momentum $l$ of $^{10}$Be+n. See the text for further details.}
\label{fig:Be11_cluster_formfactor}
\end{figure}
An insight into the wave functions of the two bound states of $^{11}$Be is provided in Fig.~\ref{fig:Be11_cluster_formfactor}. Cluster form factors calculated according to Eq.~(\ref{eq:NCSMC_formf}) using the N$^2$LO$_{\rm SAT}$ interaction are presented for the $1/2^+$ ground state (left panel) and the first excited $1/2^-$ state (right panel). A clearly extended halo structure beyond 20 fm can be identified for the $S$ wave and $P$ wave of the $^{10}$Be$(0^+)$+n relative motion for the $1/2^+$ and $1/2^-$ states, respectively. The phenomenological energy adjustment to reproduce the $^{10}$Be+n experimental separation energies, the NCSMC-pheno approach~\cite{Eraly2016}, only slightly influences the asymptotic behavior of the $S$ wave and $P$ wave, as seen by comparing the solid and dashed black curves, while other partial waves are even indistinguishable on the plot resolution. The corresponding $1/2^+$ g.s. spectroscopic factors for the NCSMC-pheno approach, obtained by integrating the squared cluster form factors in the left panel of Fig.~\ref{fig:Be11_cluster_formfactor}, are $S{=}0.90$ ($S$ wave) and $S{=}0.16$ ($D$ wave). The S-wave asymptotic normalization coefficient (ANC) is 0.786 fm$^{-1/2}$. In Table~\ref{tab:11Be}, we summarize the ANCs and spectroscopic factors of the two bound states and compare them with ANC values from knockout reaction~\cite{Hebborn2021} and transfer reaction~\cite{Yang2018} analyses. The NCSMC cluster form factors can be contrasted with those obtained within NCSM, computed as discussed below Eq.~(\ref{eq:NCSMC_formf}), shown by dotted lines in Fig.~\ref{fig:Be11_cluster_formfactor}. They drop off to zero at $\approx 8$ fm. Interestingly, the NCSM spectroscopic factors are comparable to the NCSMC ones.
%
\begin{table}[H]
\centering
\renewcommand{\arraystretch}{1.33}
\setlength{\tabcolsep}{10pt}
\begin{tabular}{l c c c}
    \rowcolor{blue!100!gray!10}
    ${}^{11}\mathrm{Be}$($1/2^+$) & ANC$^2$ $[ \mathrm{fm}^{-1} ]$ & $S$-wave spectr. factor & $D$-wave spectr. factor  \\
    \hline
    NCSMC-pheno & 0.618  & 0.90 & 0.16   \\
    \hline
     Hebborn~\cite{Hebborn2021} & $0.62 \pm 0.06 \pm 0.09$ &  &   \\
    \hline
    Yang~\cite{Yang2018} & $0.616 \pm 0.001$ & & \\
    \hline
    \rowcolor{blue!100!gray!10}
    ${}^{11}\mathrm{Be}$($1/2^-$) & ANC $[ \mathrm{fm}^{-1/2} ]$ & \multicolumn{2}{l}{$^{10}$Be($0^+$)+n $P$-wave spectr. factor} \\
    \hline
    NCSMC-pheno & 0.129  & 0.85  &    \\
    \hline
    Yang~\cite{Yang2018} & $0.135\pm 0.005$ & & \\
    \hline
\end{tabular}
\vspace{0.15cm}\caption{\label{tab:11Be} ANC$^2$ of the  $S$-wave neutron halo $1/2^+$ ground state of ${}^{11}\mathrm{Be}$ and the $S$- and $D$-wave spectroscopic factors and ANC of the $1/2^-$ halo excited state of $^{11}$Be in the $^{10}$Be($0^+$)+n $P$-wave channel with the corresponding spectroscopic factor. Calculations performed with the N$^2$LO$_{\rm SAT}$ interaction within the NCSMC-pheno are compared to results from Refs.~\cite{Hebborn2021,Yang2018}.}
\end{table}

It is known that using the chiral 3N interaction with a non-local regularization improves the description of nuclear radii compared to experiment~\cite{HUTHER2020135651}. Similarly, NCSMC investigations presented in this subsection suggest importance of a non-local regularization of the chiral 3N interaction for the reproduction of the parity inversion in $^{11}$Be ground state.

\subsection{Halo ground-state of $^{15}$C}\label{subsec:C15}

The $^{15}$C is a well known one-neutron halo nucleus~\cite{Tanihata1996,Riisager_2013}. With a rather small one-neutron separation energy of 1.22 MeV, its ground state can be well described as $^{14}$C in its $0^+$ ground state and a loosely bound neutron in the $1s_{1/2}$ orbital. It is relevant for nuclear astrophysics. The $^{15}$C synthesis through one-neutron radiative capture, $^{14}$C(n,$\gamma$)$^{15}$C, has been suggested to be part of neutron-induced CNO cycles, which take place in the helium-burning zone of asymptotic-giant-branch (AGB) stars~\cite{Wiescher_1999}. This reaction is also the doorstep to the production of heavy elements in inhomogeneous big-bang nucleosynthesis~\cite{Kajino1990} and it has been shown to be part of possible reaction routes in the nuclear chart during the $r$ process in Type II supernovae~\cite{Terasawa_2001}. This cross section is also important as a benchmark both for theories and experiments as it can be measured directly and used for validation of the Coulomb breakup method for the neutron capture cross section determination using the $^{15}$C beam~\cite{Moschini2019}.

%
\begin{figure}[H]
\centering
\subfloat[\centering]{\includegraphics[width=8.2cm]{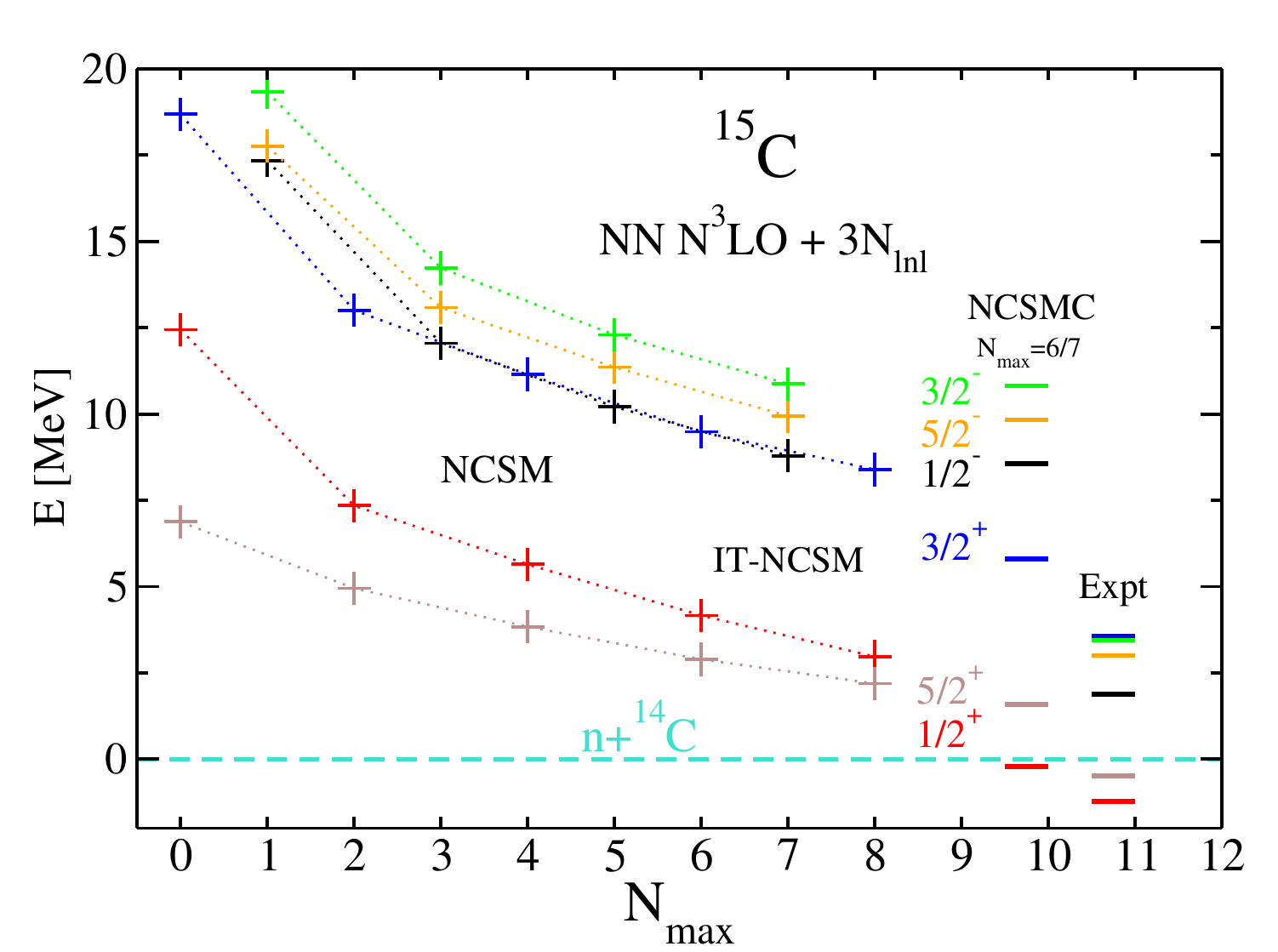}}
\subfloat[\centering]{\includegraphics[width=8.2cm]{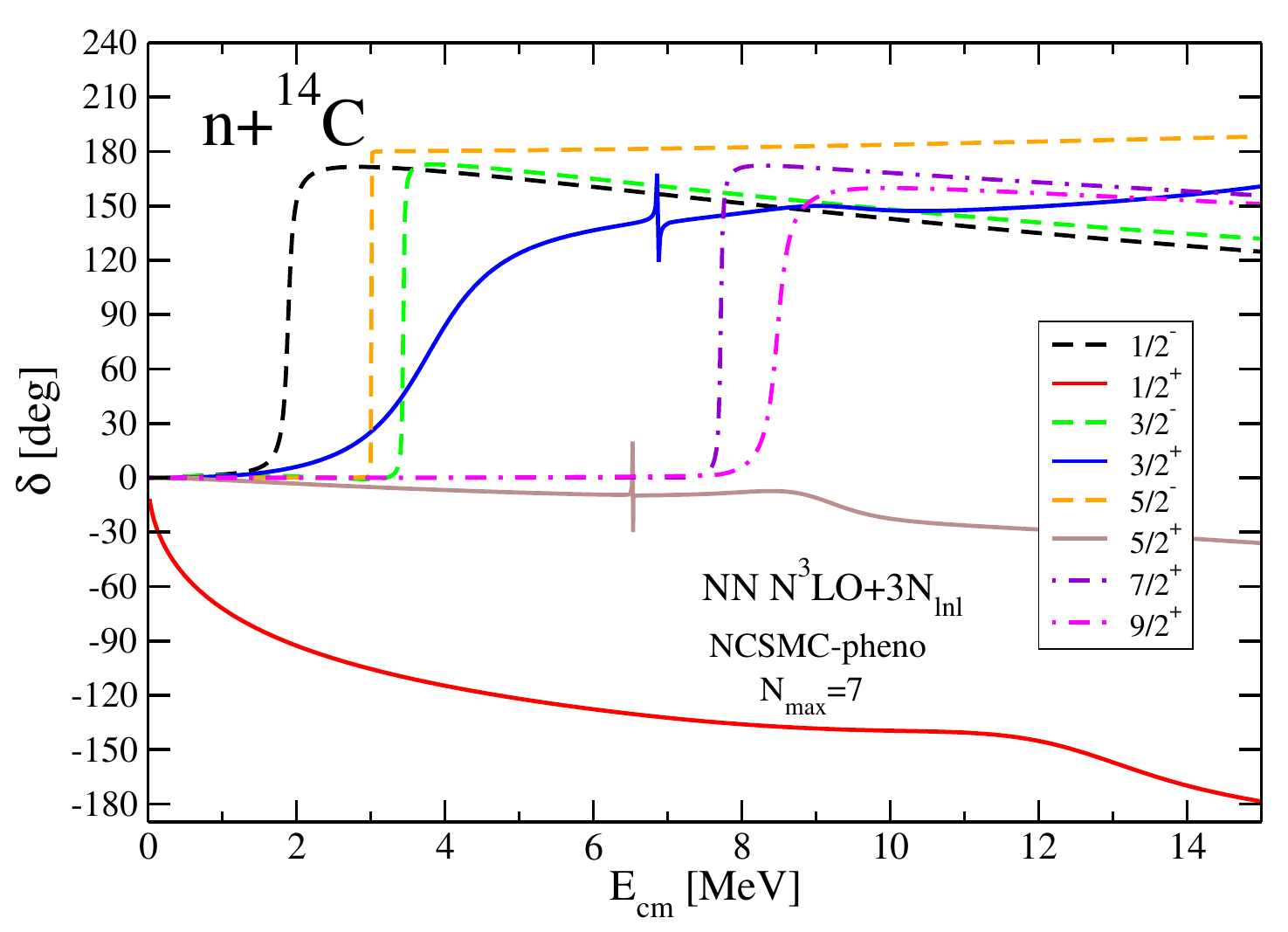}}
\caption{(a) Calculated energies of low-lying states of $^{15}$C compared to experiment. The crosses correspond to NCSM calculations in basis spaces up to $N_{\rm max}{=}8$. The NCSMC calculations has been performed in $N_{\rm max}{=}7$. All energies are with respect to the $^{14}$C+n threshold, the calculated one obtained in the consistent $N_{\rm max}$ space. (b) Diagonal $^{14}$C+n phase shift dependence on the energy in the center of mass obtained within the NCSMC-pheno approach in $N_{\rm max}{=}7$ space. The NN N$^3$LO + 3N$_{\rm lnl}$ interaction and the HO frequency of $\hbar\Omega{=}20$ MeV has been used in all calculations. See the text for further details.}
\label{fig:C15_energies_phaseshifts}
\end{figure}
We have investigated $^{15}$C within the NCSMC using the NN N$^3$LO+3N$_{\rm lnl}$ interaction~\cite{Soma2000} that was SRG evolved using $\lambda_{\rm SRG}{=}2$ fm$^{-1}$ with the 3N induced terms included. In the basis expansion~(\ref{ncsmc_wf}), we considered the $^{14}$C+n cluster including the $^{14}$C $0^+$ ground state and the first-excited $2^+$ state as well as 7(3) positive(negative)-parity NCSM eigenstates of $^{15}$C. The underlying NCSM calculations have been performed up to $N_{\rm max}{=}8$ and the NCSMC in $N_{\rm max}{=}7$, see Fig.~\ref{fig:C15_energies_phaseshifts} (a) where energies of low-lying states of both parities are presented. The full basis space has been used up to $N_{\rm max}{=}5$(6) for $^{15}$C ($^{14}$C) while the importance truncated (IT) NCSM~\cite{Roth2007,Roth2009,Kruse2013} has been applied for higher $N_{\rm max}$. One can see the significant increase of binding for positive parity states in NCSMC compared to NCSM. The negative-parity states are, on the other hand, basically unchanged when continuum is taken into account. The $1/2^+$ ground state is bound with respect to the $^{14}$C+n threshold in NCSMC although less than in experiment. In contrast, in NCSM the $5/2^+$ state is predicted to be the ground state up to at least $N_{\rm max}{=}8$. It gains binding in NCSMC although it remains unbound in $N_{\rm max}{=}7$ contrary to experiment. 

To gain insight into the structure of the experimentally bound $1/2^+$ and $5/2^+$ states and to facilitate the calculation of the $^{14}$C(n,$\gamma$)$^{15}$C cross section, we apply the NCSMC-pheno approach~\cite{Eraly2016}. We adjust the NCSM calculated excitation energy of the $^{14}$C $2^+$ state from $N_{\rm max}{=}6$ calculated 8.73 MeV to experimental 7.01 MeV and shift the $^{15}$C NCSM eigenstates to reproduce experimental energies of the bound states (and thus the experimental threshold) and known low-lying resonances in the NCSMC calculations. As a result, we obtain the $1/2^+$ and $5/2^+$ states bound as in experiment and predict the phase shifts shown in Fig.~\ref{fig:C15_energies_phaseshifts} (b) with the broad $3/2^+$ resonance and the very narrow $1/2^-, 5/2^-, 3/2^-$ resonances matching experimental centroids. At the same time, we predict narrow $5/2^+, 3/2^+, 7/2^+, 9/2^+$ resonances close to the $^{14}$C($2^+$)+n threshold. 

%
\begin{figure}[H]
\centering
\subfloat[\centering]{\includegraphics[width=8.4cm]{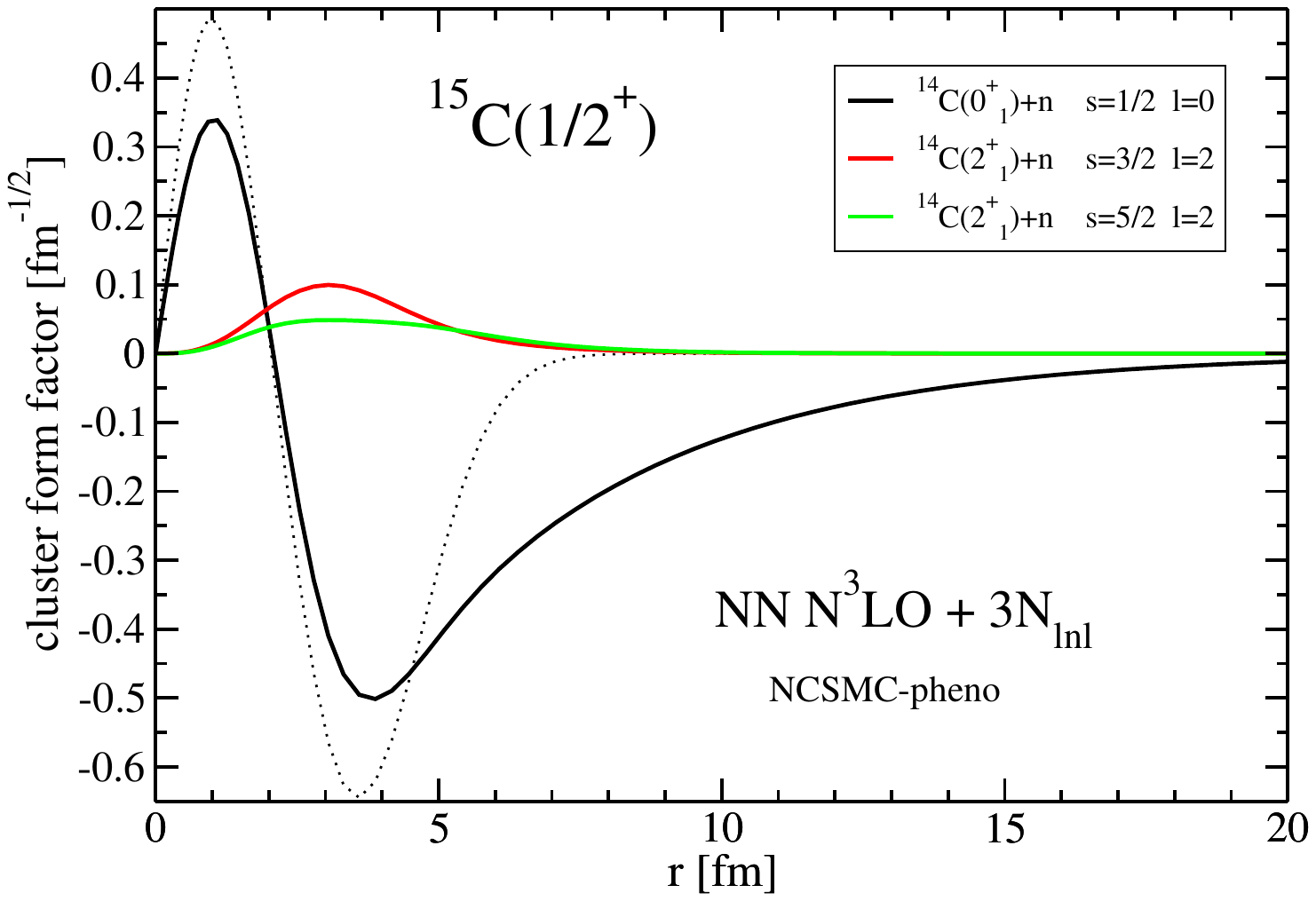}}
\subfloat[\centering]{\includegraphics[width=8.4cm]{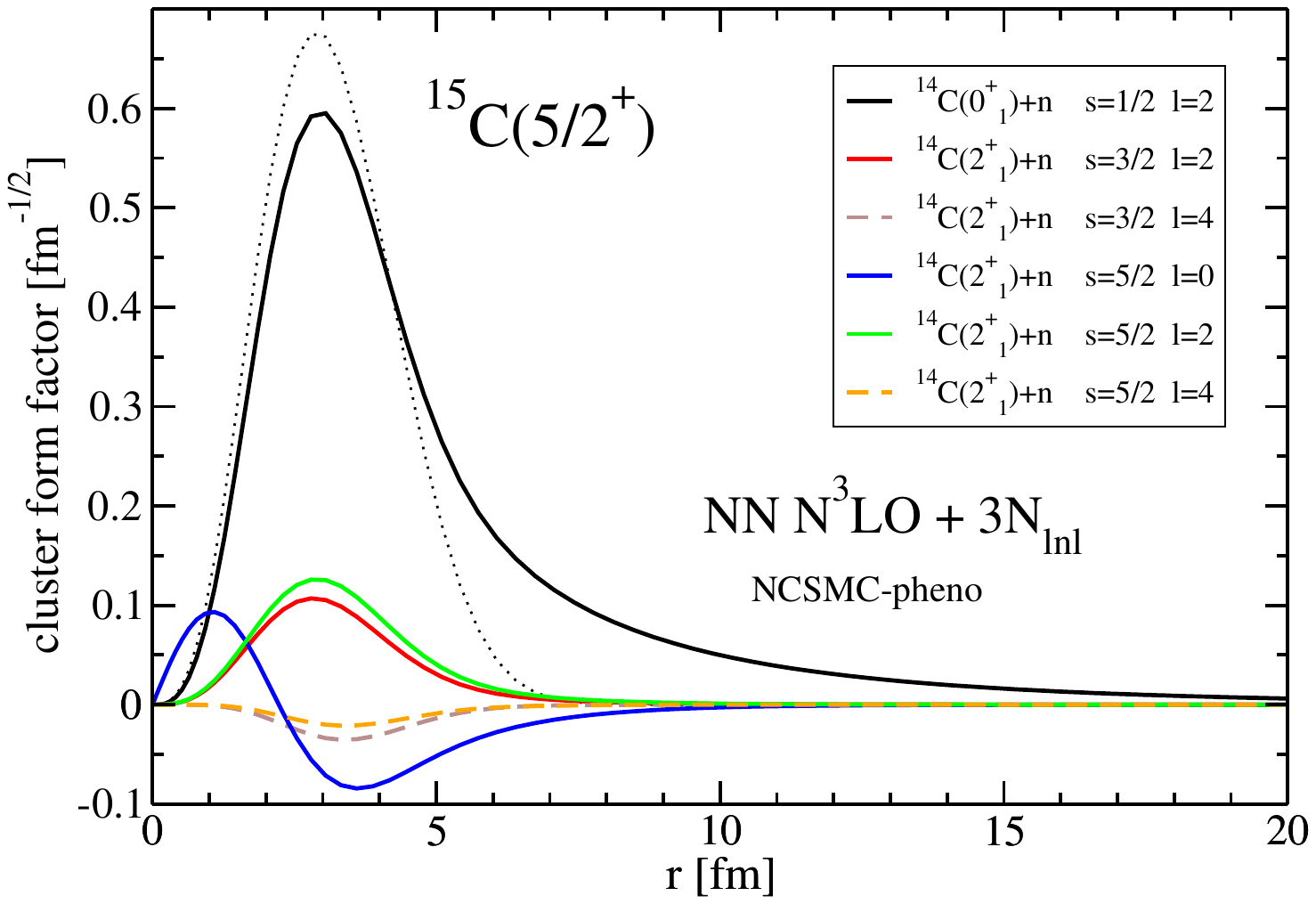}}
\caption{Cluster form factors of $^{15}$C $1/2^+$ (a) and $5/2^+$ (b) states obtained with the NN N$^3$LO+3N$_{\rm lnl}$ interaction. The solid lines show the NCSMC-pheno results, the black dotted lines are the NCSM $S$-wave (a) and $D$-wave (b) results for the $^{14}$C in the $0^+_1$ ground state. See the text for further details.}
\label{fig:C15_cluster_formfactor}
\end{figure}
In Fig.~\ref{fig:C15_cluster_formfactor}, we show the cluster form factors of the two $^{15}$C bound states, the $1/2^+$ ground state (a) and the $5/2^+$ excited state (b) obtained within the NCSMC-pheno approach. The $1/2^+$ state clearly manifests an $S$-wave neutron halo extending beyond 20 fm. The calculated ANC $C_{1/2^+}{=}1.282$ fm$^{-1/2}$ ($C^2_{1/2^+}{=}1.644$ fm$^{-1}$) is in an excellent agreement with that of $C^2_{1/2^+}{=}1.59\pm0.06$ fm$^{-1}$ obtained in Ref.~\cite{Moschini2019} using the halo EFT analysis of the $^{14}$C(d,p)$^{15}$C transfer reaction. Similarly, it is in agreement with the ANC determination of $C^2_{1/2^+}{=}1.57\pm0.30\pm0.18$ fm$^{-1}$ from $^{15}$C one-neutron knockout reaction data analysis~\cite{Hebborn2021}. The corresponding spectroscopic factor obtained by integrating the square of the cluster form factor is 0.96. $D$-wave contributions from the $^{14}$C($2^+$)+n are quite small.

The $5/2^+$ state is dominated by $^{14}$C($0^+$) and neutron in a $D$ wave. Given its weak binding of just 0.48 MeV the cluster form factor extends beyond 15 fm. The NCSMC-pheno calculated ANC is $C_{5/2^+}{=}0.048$ fm$^{-1/2}$ in a good agreement with the the value of $0.0595(36)$ fm$^{-1/2}$ determined from the analysis of the $^{14}$C(d,p)$^{15}$C transfer reaction~\cite{Mukhamedzhanov2011}. The corresponding spectroscopic factor is 0.90. There are also small but visible contributions by $^{14}$C($2^+$) and neutron in $S$ and $D$ waves while the $G$ waves are negligible.

The dotted lines in Fig.~\ref{fig:C15_cluster_formfactor} show the corresponding cluster form factors obtained within NCSM. While their shapes and the spectroscopic factors are similar to the NCSMC ones, their extent is drastically different as they become negligible beyond ${\sim}7$ fm. 

%
\begin{figure}[H]
\includegraphics[width=0.7\textwidth]{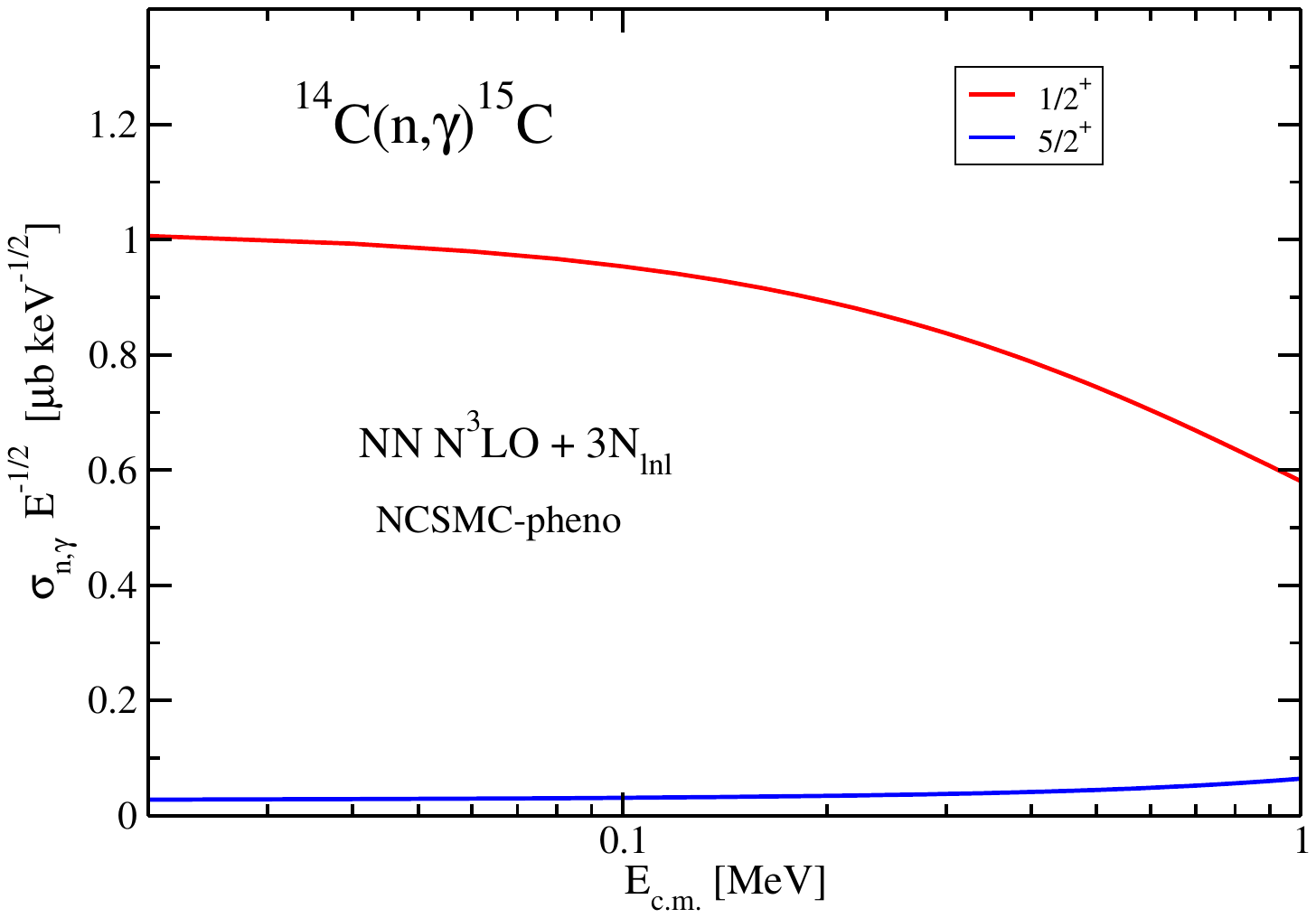}
\caption{Cross sections for the radiative capture $^{14}$C(n,$\gamma$)$^{15}$C to the $1/2^+$ and $5/2^+$ final states obtained within the NCSMC-pheno approach using the NN N$^3$LO + 3N$_{\rm lnl}$ interaction.}
\label{fig:14Cngamma15C}
\end{figure}  
Applying the NCSMC-pheno calculations discussed above, we have computed the cross section of the $^{14}$C(n,$\gamma$)$^{15}$C radiative capture reaction. The energy-scaled cross section for energy up to 1 MeV is displayed in Fig.~\ref{fig:14Cngamma15C} showing separately the capture to the ground state and to the $5/2^+$ excited state. Overall, the shape and magnitude is in line with recent experimental determinations~\cite{Reifarth2008}. Following the pioneering measurement~\cite{Beer1992}, it is customary to compare theoretical and experimental cross sections at $E_{\rm c.m.}{=}23.3$ keV. At this energy, we have obtained in the present NCSMC-pheno calculations $\sigma_{{\rm n},\gamma}$($1/2^+$)${=}4.79\, \mu$b, $\sigma_{{\rm n},\gamma}$($5/2^+$)${=}0.13\, \mu$b, i.e., the total $\sigma_{{\rm n},\gamma}{=}4.92\, \mu$b. This is in a good agreement with the value of 4.66(14) $\mu$b reported in Ref.~\cite{Moschini2019} (for the capture to the ground state) and the result of 4.75 $\mu$b obtained in Ref.~\cite{Tkachenko2025}, while it is slightly higher than the recent measurement by Jiang \textit{et al.}~\cite{Jiang_2025} reporting 3.89(76) $\mu$b.

\begin{table}[H]
\centering
\renewcommand{\arraystretch}{1.33}
\setlength{\tabcolsep}{10pt}
\begin{tabular}{l c c c}
    \rowcolor{blue!100!gray!10}
    ${}^{15}\mathrm{C}$($1/2^+$) $S$ wave & ANC $ [ \mathrm{fm}^{-1/2} ]$ & Spectr. factor & \\
    \hline
    NCSMC-pheno & 1.282 & 0.96 &  \\
    \hline
     Moschini ~\cite{Moschini2019} & 1.26(2)  & 1  &  \\
    \hline
     Hebborn~\cite{Hebborn2021} & 1.25(12)   & 1  &   \\
    \hline
     Jiang~\cite{Jiang_2025} & 1.16(15) & 0.68(14)  & \\
    \hline
    \rowcolor{blue!100!gray!10}
    ${}^{15}\mathrm{C}$($5/2^+$) $D$ wave & ANC $ [ \mathrm{fm}^{-1/2} ]$ & Spectr. factor & \\
    \hline
     NCSMC-pheno & 0.048 & 0.90 & \\
    \hline
     Mukhamedzhanov~\cite{Mukhamedzhanov2011} & 0.0595(36) & 1 & \\
    \hline
    \rowcolor{blue!100!gray!10}
     $\sigma_{{\rm n},\gamma}$ [$\mu$b] at $E_{\rm c.m.}{=}23.3$ keV& $1/2^+$  & $5/2^+$ & Total   \\
    \hline
    NCSMC-pheno & 4.79 & 0.13 & 4.92 \\
    \hline
     Moschini~\cite{Moschini2019} & 4.66(14) &  &  \\
     \hline
     Tkachenko~\cite{Tkachenko2025} & & & 4.75 \\
     \hline
     Jiang~\cite{Jiang_2025} & 3.89(76) & & \\
    \hline \\
\end{tabular}
\vspace{0.15cm}\caption{\label{tab:15C} NCSMC-pheno calculated ANCs and spectroscopic factors for the $ 1/2^{+} $ and $5/2^+$ bound states of $ {}^{15}\mathrm{C} $ and the $^{14}$C(n,$\gamma$)$^{15}$C cross section at $E_{\rm c.m.}{=}23.3$ keV for the capture to the $1/2^+$ and $5/2^+$ final states compared to results obtained by other methods.}
\end{table}
In Table~\ref{tab:15C}, we summarize the presently obtained ANCs, spectroscopic factors and cross sections and compare them to results in literature. For a more in depth review of $^{14}$C(n,$\gamma$)$^{15}$C cross section data, we refer the reader to Ref.~\cite{Tkachenko2025}.

\subsection{P-wave halo nucleus $^8$B}\label{subsec:B8}

$^8$B plays an important role in astrophysics as neutrinos from its $\beta$ decay form the higher-energy part of the solar neutrino flux. It is produced in the solar proton-proton reaction chain through the proton radiative capture on $^7$Be, the $^7$Be(p,$\gamma$)$^8$B reaction. The $2^+$ ground state of $^8$B is bound by only ${\approx}137$ keV with respect to the $^7$Be+p threshold. Consequently, it is anticipated that it manifests a $P$-wave proton halo. The $^7$Be(p,$\gamma$)$^8$B capture reaction and the $^8$B structure have been investigated recently within the NCSMC approach using several sets of chiral NN+3N interactions. We focus here on the results obtained with the NN-N$^4$LO~\cite{Entem2017}+3N$^*_{\rm lnl}$~\cite{Kravvaris2023,Jokiniemi2024,Girlanda2011} chiral interaction (denoted as NN N$^4$LO + 3N$_\mathrm{lnlE7}$ in some of the figures). In this interaction, an additional sub-leading contact term ($E_7$) enhancing the spin-orbit strength~\cite{Girlanda2011} has been introduced to the 3N force. It appears to be the best performing interaction available to us currently. Again, the SRG evolution was applied with 3N induced terms included.

\begin{figure}[H]
\includegraphics[width=0.7\textwidth]{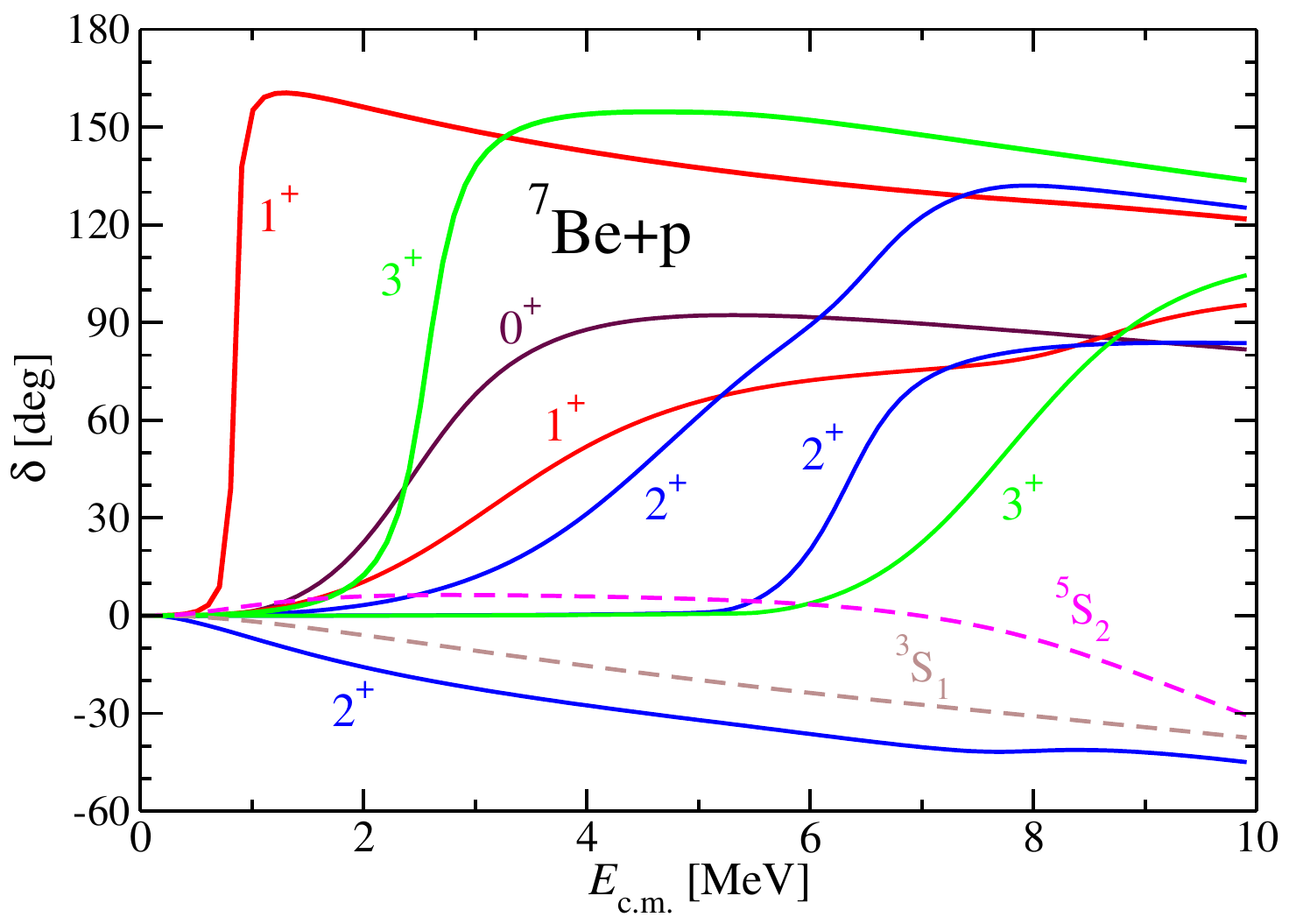}
\caption{$^7$Be+p eigenphase shifts (solid lines) and $^3S_1$ and $^5S_2$ diagonal phase shifts (dashed lines) obtained from the NCSMC approach with the NN-N$^4$LO+3N$^{*}_{\rm lnl}$ interaction. Figure from Ref.~\cite{Kravvaris2023}.}
\label{fig:pBe7_phaseshift}
\end{figure}  
The positive parity eigenphase shifts for $^7$Be+p scattering obtained using the NN-N$^4$LO+3N$^*_{\rm lnl}$ chiral interaction, presented in Fig.~\ref{fig:pBe7_phaseshift}, show the well-established $1^+$ and $3^+$ $^8$B resonances as well as several other, yet unobserved, broad resonances. The NCSMC $S$-wave phase shifts manifest scattering length signs consistent with those determined in recent measurements (negative for $^5S_2$, positive for $^3S_1$~\cite{Paneru2019}). We find it is difficult to produce a bound $^8$B ground state with this as well as with other chiral interactions~\cite{Kravvaris2023}. Rather, we obtained a very narrow near-threshold $2^+$ resonance that is not visible in the figure. 

To investigate the properties of the weakly-bound $^8$B $2^+$ ground state, we resort to the NCSMC-pheno approach~\cite{Eraly2016}, i.e., with the $^7$Be(g.s.)+p separation energy adjusted to the experimental value of 137 keV. This is achieved by shifting the NCSM eigenenergies of $^7$Be so that the excitation energies (and therefore thresholds) match the experimental ones exactly. Furthermore, $^8$B NCSM eigenenergies in the $2^+$ channel are also modified bringing the NCSMC states in the experimentally observed positions.

\begin{figure}[H]
\includegraphics[width=0.8\textwidth]{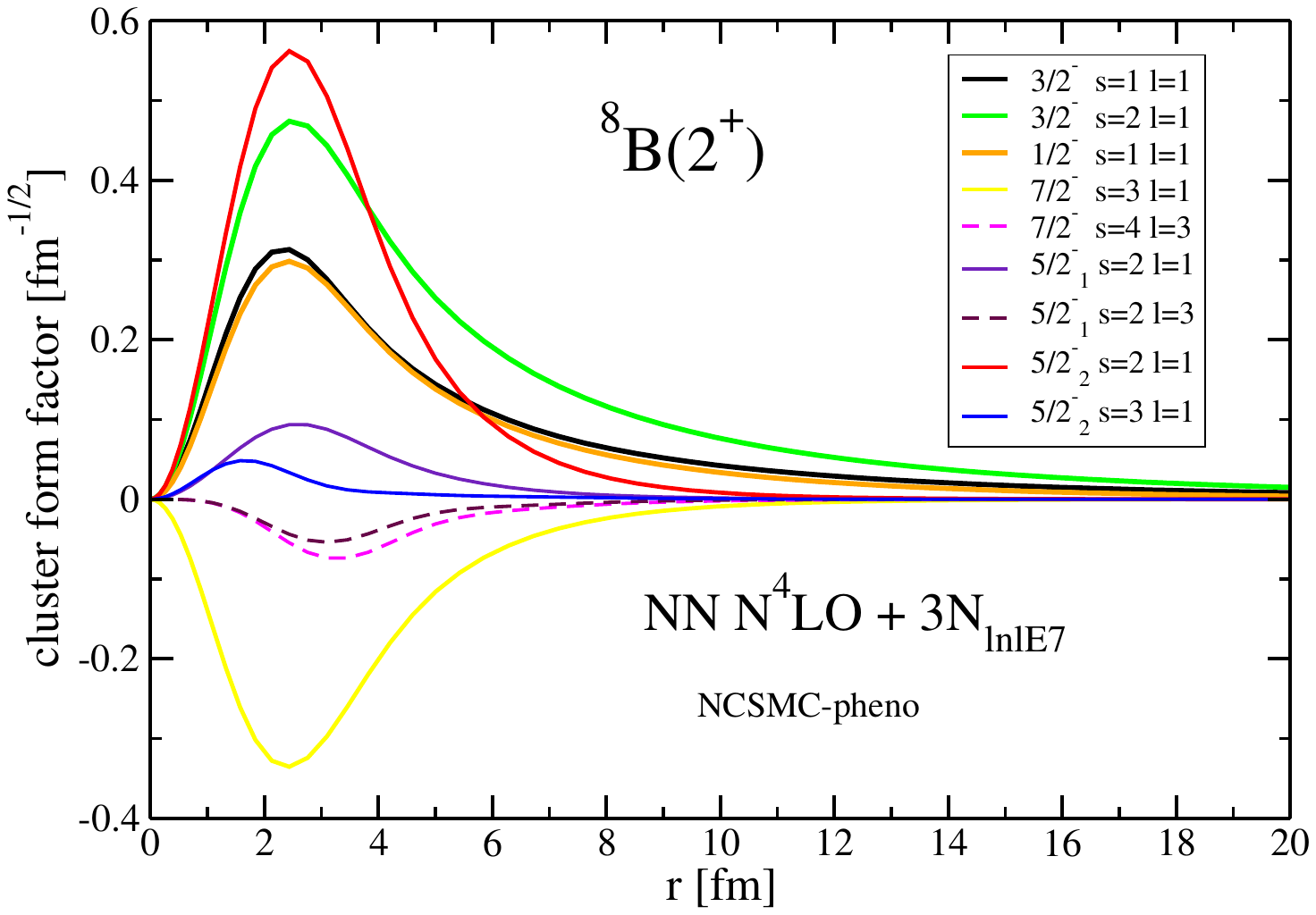}
\caption{Cluster form factors of $^{8}$B $2^+$ ground state obtained with the NN-N$^4$LO+3N$^*_{\rm lnl}$ interaction within the NCSMC-pheno. The legend columns show the $^7$Be eigenstate, the channel spin $s$ and the relative orbital momentum $l$ of $^7$Be+p. See the text for further details.}
\label{fig:B8_formfactor}
\end{figure}
In Fig.~\ref{fig:B8_formfactor}, we present the cluster form factor for the $2^+$ ground state of $^8$B obtained using the NN-N$^4$LO+3N$^*_{\rm lnl}$ interaction within the NCSMC-pheno approach. The dominant component is clearly the channel-spin $s{=}2$ $P$ wave of the $^7$Be(g.s.)+p that extends to a distance far beyond the plotted range. The alternative channel spin coupling, $s{=}1$, $P$ wave is less pronounced but it extends in a similar way. Of a comparable size is the $^7$Be($1/2^-$)+p $P$ wave. Remarkably, we notice a substantial contribution from the $^7$Be($5/2^-_2$)+p $P$ wave in the channel spin $s{=}2$. The other possible $s{=}3$ $P$-wave configuration is negligible. At the same time, the $^7$Be $5/2^-_2$ state is dominated by a $^6$Li+p channel-spin $s{=}3/2$ $P$-wave configuration. Within the NCSM framework relevant to the present calculations this was shown (for the mirror $^7$Li+n system) in ref.~\cite{Navratil2004cl}. Therefore, such a large contribution of the $s{=}2$ $^7$Be($5/2^-_2$)+p $P$ wave to the $^8$B ground state seems to indicate the presence of two antiparallel protons outside of a $^6$Li core, and that their exchanges are important. Clearly, for a realistic description of the $^8$B ground state, this state must be taken into account. Finally, we note that the $^7$Be($7/2^-$)+p $P$-wave component is also substantial. The calculated ANCs, $C_{p1/2}{=}0.34$ fm$^{-1/2}$ and $C_{p3/2}{=}0.62$ fm$^{-1/2}$ are close to experimental values reported in Ref.~\cite{Paneru2019}, see Table~\ref{tab:8B}. 
%
\begin{table}[H]
\centering
\renewcommand{\arraystretch}{1.33}
\setlength{\tabcolsep}{10pt}
\begin{tabular}{l c c}
    \rowcolor{blue!100!gray!10}
    ${}^{8}\mathrm{B}$($2^+$) & $C_{p1/2}$   & $C_{p3/2}$  \\
    \hline
    NCSMC-pheno & 0.34(1) & 0.62(2)   \\
    \hline
     Paneru~\cite{Paneru2019} & 0.315(9) & 0.66(2)  \\
    \hline
\end{tabular}
\vspace{0.15cm}\caption{\label{tab:8B} ANCs, in $[ \mathrm{fm}^{-1/2} ]$, of the  $P$-wave proton halo $2^+$ ground state of ${}^{8}\mathrm{B}$ obtained with the NN-N$^4$LO+3N$^*_{\rm lnl}$ interaction within the NCSMC-pheno compared to experimental results~\cite{Paneru2019}. Estimates of chiral truncation errors are shown for theoretical values.}
\end{table}
%


\subsection{Excited halo states in $^{10}$Be}\label{subsec:Be10}

\begin{figure}[t]
    \centering
    \includegraphics[width=\linewidth]{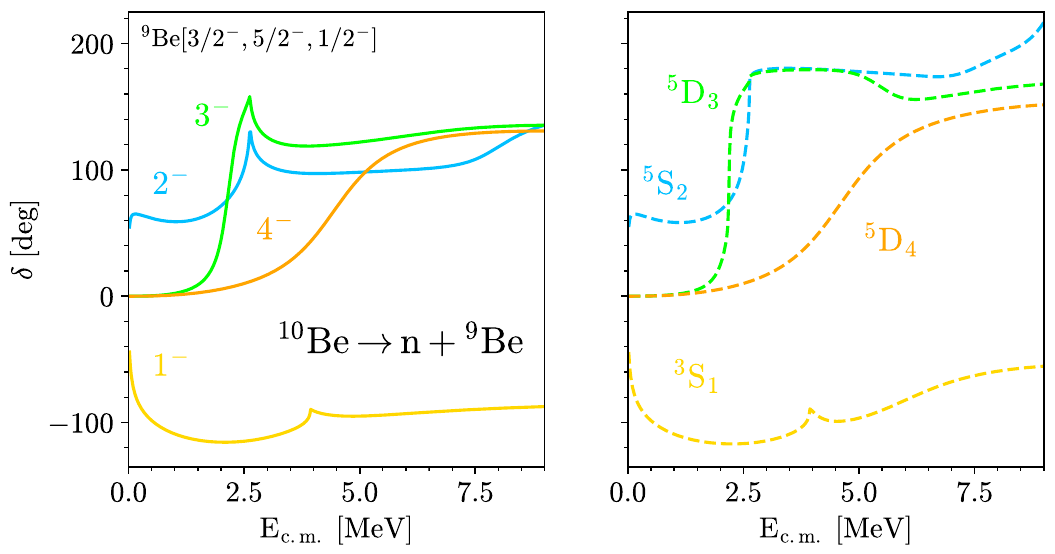}
    \caption{${}^{10}\mathrm{Be} \rightarrow \mathrm{n} + {}^{9}\mathrm{Be}$ negative parity eigenphase (left) and phase (right) shifts obtained from the NCSMC approach with the NN-N$^4$LO+3N$_{\rm lnl}$ chiral interaction and an $ N_{\mathrm{max}} = 9 $ model space. The $ {3/2}^{-} $, $ {5/2}^{-} $ and $ {1/2}^{-} $ configurations are included in the ${}^{9}\mathrm{Be}$ mass partition.}
    \label{section:10Be:figure:negativeparity_eigenphase}
\end{figure}

The lowest-lying part of the dense spectrum in $ {}^{10}\mathrm{Be} $ is well understood. What remains however is a clear understanding of the excited $ J^{\pi} = 1^{-}, \, 2^{-} $ states for which there exists inherent nuclear structure interest. These two states have been experimentally measured close to the $ \mathrm{n} + {}^{9}\mathrm{Be} $ threshold, and both are thus anticipated to have some kind of exotic structure; either strong clustering or $ S $-wave halo formation~\cite{PhysRevC.74.034312}. The appearance of two exotic, possibly-$ S $-wave halo states so close together is a rather unique feature of the $ {}^{10}\mathrm{Be} $ spectrum, the classification of which have naturally garnered interest over several decades, partially due to the inconclusive literature on the matter. Recently, there has been novel experimental effort in classifying the nature of these states; an experiment from TRIUMF ISAC-II investigating the angular distributions of both states via the $ {}^{11}\mathrm{Be}( \, \mathrm{p}, \, \mathrm{d} \,) $ transfer reaction provided inconclusive information regarding the cluster vs. halo nature of the states~\cite{PhysRevC.104.044601}, while a further proposal to examine the structure of the $ 2^{-} $ via the one-neutron-transfer reaction $ {}^{9}\mathrm{Be} \big( \, {}^{11}\mathrm{Be}, \, {}^{10}\mathrm{Be}^*[2^-] \, \big) $ is currently in preparation at ISOLDE~\cite{chen2023probing}. In addition to these exotic bound states, there exists interest in the $ J^{\pi} = 3^{-} $ resonance state which sits slightly above the $ \mathrm{n} + {}^{9}\mathrm{Be} $ threshold. As the isospin mirror of $ {}^{10}\mathrm{C} $, characterization of these excited states and resonances can provide insight into isospin symmetry breaking effects arising from the strong and electromagnetic sectors.

\begin{figure}[t]
    \centering
    \includegraphics[width=\linewidth]{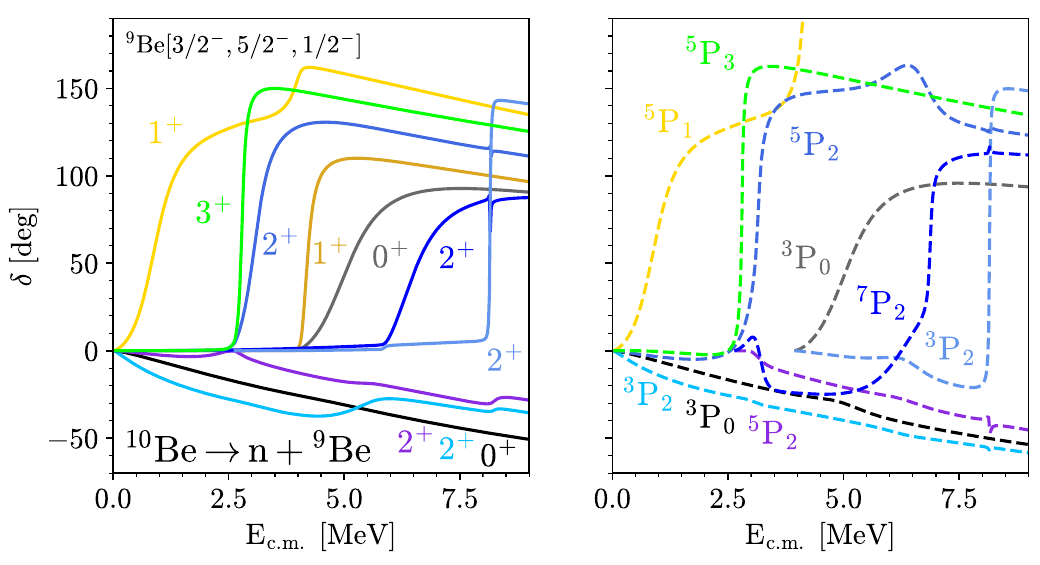}
    \caption{${}^{10}\mathrm{Be} \rightarrow \mathrm{n} + {}^{9}\mathrm{Be}$ positive parity eigenphase (left) and phase (right) shifts obtained from the NCSMC approach with the NN-N$^4$LO+3N$_{\rm lnl}$ chiral interaction and an $ N_{\mathrm{max}} = 9 $ model space. The $ {3/2}^{-} $, $ {5/2}^{-} $ and $ {1/2}^{-} $ configurations are included in the ${}^{9}\mathrm{Be}$ mass partition.}
    \label{section:10Be:figure:positiveparity_eigenphase}
\end{figure}

In Figs.~\ref{section:10Be:figure:negativeparity_eigenphase} and~\ref{section:10Be:figure:positiveparity_eigenphase}, we present the NCSMC predictions for the negative and positive parity phase shifts, respectively, for the $\mathrm{n} + {}^{9}\mathrm{Be}$ scattering process with the eigenphase shifts shown on the left and phase shifts shown on the right. These are obtained with the NN-N$^4$LO+3N$_{\rm lnl}$ chiral interaction and within an $ N_{\mathrm{max}} = 8, \, 9 $ model space depending on the parity. We identify the total spin parity with a given eigenphase shift and the partial wave channel for a given phase shift in the same color as the curves. We use a mass partition including the $ {3/2}^{-} $, $ {5/2}^{-} $ and $ {1/2}^{-} $ states of $ {}^{9}\mathrm{Be} $ and, with this configuration of states, the NCSMC binds four out of the six (both positive and negative parity) experimentally observed bound states~\cite{TUNL.Be10.2004}. We notably miss the excited $ J^{\pi} = 0^{+} $ state due to lacking consideration of alpha clustering in the partitions, which has been known to contribute significantly to the structure for some time~\cite{PRC.59.817.1999, CPL.27.022103.2010}.

In Fig.~\ref{section:10Be:figure:negativeparity_eigenphase}, dominated by the $ {}^{3}\mathrm{S}_{1} $ partial wave, we find that the $ 1^{-} $ state is readily bound in the NCSMC calculation, albeit sitting shallow at $ -0.0295 \ \mathrm{MeV} $ with respect to the already-quite-shallow experimental observation of $ -0.8523 \ \mathrm{MeV}$. On the other hand, the $ 2^{-} $ remains unbound in the current model space, but exhibits the expected near-threshold behavior for resonance conversion to a bound state. It is possible that an increase in the model space dimension would drive this state below threshold, but it could be that mixing between additional mass partitions are necessary to describe this state, e.g., ${}^{6}\mathrm{He}+\alpha $. Moreover, predicated on the inclusion of the full set of low-lying $ {}^{9}\mathrm{Be} $ states which could contribute to the formation of these structures, the observed closeness of these states in the spectrum is driven by their sharing of the same underlying $ {}^{9}\mathrm{Be} [ {3/2}^{-} ]$ configuration with the neutron spin anti-aligned (${}^{3}\mathrm{S}_{1}$) or aligned (${}^{5}\mathrm{S}_{2}$) with the nuclear spin -- but always with $ l = 0 $ relative orbital momentum. The inclusion of the $ {5/2}^{-} $ and $ {1/2}^{-} $ prove to be of little consequence to the formation of these states, with stability in the phase shifts as each consecutive state is added. A further improvement to our description of the $ 1^{-} $ would likely come from inclusion of the $ {}^{6}\mathrm{He} \, + \, \alpha $ mass partition, as analysis via microscopic cluster models with a partition of $ \alpha \, + \, \alpha \, + \, n \, + \, n $ suggests that a significant contribution to the structure of these states is driven by $ \alpha $-clustering effects~\cite{PTEP.2020.023D02.2020}; notably the $ 2^{-} $ is not bound in such an approach, suggesting reduced effects of $ \alpha $-clustering on the structure of said state.

As expected, we further see the appearance of the known $ 3^{-} $ resonance coming from the $ {}^{5}\mathrm{D}_{2} $ partial wave channel, consistent with that which has been similarly seen in calculations of $ {}^{10}\mathrm{C} $ with the NCSMC~\cite{gennari2021ab}. Lastly, a broad $ 4^{-} $ resonance from the $ {}^{5}\mathrm{D}_{4} $ channel appears at relatively low c.m. energies, corresponding to the expected resonance at $ 2.46 \ \mathrm{MeV} $. While the partial waves which dominate the contribution to these states have higher total momentum, it comes primarily from the relative orbital momentum of $ l = 2 $ between the neutron and $ {}^{9}\mathrm{Be} $ cluster, i.e., they are still largely built upon the $ {3/2}^{-} $ of $ {}^{9}\mathrm{Be} $ with neutron spin either anti-aligned or aligned with the nuclear spin.


\begin{table}[t]
\centering
\renewcommand{\arraystretch}{1.33}
\setlength{\tabcolsep}{10pt}
\begin{tabular}{c c c c c}
    \hline
    \rowcolor{blue!100!gray!10}
    State of $ {}^{9}\mathrm{Be} $ & $ l $ & $ S $ & ANC $ [ \mathrm{fm}^{-1/2} ]$ & ANC Pheno $ [ \mathrm{fm}^{-1/2} ]$ \\
    \hline
    $3/2^{-}$ & 0 & 1 & 0.363 & 0.951 \\
    \hline
    $3/2^{-}$ & 2 & 1 & $0.7\times10^{-3}$ & $0.392\times10^{-1}$ \\
    \hline
    $3/2^{-}$ & 2 & 2 & $-0.244\times10^{-3}$ & $-0.137\times10^{-1}$ \\
    \hline
    \rowcolor{blue!100!gray!10}
     &  &  &  &  \\
    \hline
    $5/2^{-}$ & 2 & 2 & $-0.102$ & $-0.230$ \\
    \hline
    $5/2^{-}$ & 2 & 3 & $0.104\times10^{-1}$ & $0.399\times10^{-1}$ \\
    \hline
    $5/2^{-}$ & 4 & 3 & $-0.603\times10^{-4}$ & $-3.60\times10^{-3}$ \\
    \hline
    \rowcolor{blue!100!gray!10}
     &  &  &  &  \\
    \hline
    $1/2^{-}$ & 0 & 1 & 0.257 & 0.425 \\
    \hline
    $1/2^{-}$ & 2 & 1 & $0.184\times10^{-1}$ & $0.506\times10^{-1}$ \\
    \hline \\
\end{tabular}
\caption{\label{section:10Be:table:ANCs_for_1m} Asymptotic normalization coefficients (ANCs) for the $ 1^{-} $ halo state of $ {}^{10}\mathrm{Be} $, calculated with the same interaction and configuration space as described in Fig.~\ref{section:10Be:figure:negativeparity_eigenphase}. The ANCs are presented for a given spin-parity state of $ {}^{9}\mathrm{Be} $, relative orbital momentum of the neutron with respect to the $ {}^{9}\mathrm{Be} $ cluster, and total spin of the $ \mathrm{n} + {}^{9}\mathrm{Be} $ partition. The first column contains the raw ANC from the calculation, whereas the second column contains the ANCs obtained by phenomenological adjustment of the NCSMC eigenenergies to their experimental values.}
\end{table}


\begin{table}[t]
\centering
\renewcommand{\arraystretch}{1.33}
\setlength{\tabcolsep}{10pt}
\begin{tabular}{c c c c}
    \rowcolor{blue!100!gray!10}
    State of $ {}^{9}\mathrm{Be} $ & $ l $ & $ S $ & ANC Pheno $ [ \mathrm{fm}^{-1/2} ]$ \\
    \hline
    $3/2^{-}$ & 2 & 1 & $-0.288\times10^{-1}$ \\
    \hline
    $3/2^{-}$ & 0 & 2 & $-0.756$ \\
    \hline
    $3/2^{-}$ & 2 & 2 & $-0.103\times10^{-1}$ \\
    \hline
    $3/2^{-}$ & 4 & 2 & $-0.274\times10^{-4}$ \\
    \hline
    \rowcolor{blue!100!gray!10}
     &  &  &  \\
    \hline
    $5/2^{-}$ & 0 & 2 & $-0.451$ \\
    \hline
    $5/2^{-}$ & 2 & 2 & $0.164$ \\
    \hline
    $5/2^{-}$ & 4 & 2 & $0.849\times10^{-4}$ \\
    \hline
    $5/2^{-}$ & 2 & 3 & $0.126$ \\
    \hline
    $5/2^{-}$ & 4 & 3 & $-0.128\times10^{-3}$ \\
    \hline
    \rowcolor{blue!100!gray!10}
     &  &  &  \\
    \hline
    $1/2^{-}$ & 2 & 0 & $-0.184\times10^{-1}$ \\
    \hline
    $1/2^{-}$ & 2 & 1 & $-0.348\times10^{-1}$ \\
    \hline \\
\end{tabular}
\vspace{0.15cm}\caption{\label{section:10Be:table:ANCs_for_2m} Asymptotic normalization coefficients (ANCs) for the $ 2^{-} $ halo state of $ {}^{10}\mathrm{Be} $, calculated with the same interaction and configuration space as described in Fig.~\ref{section:10Be:figure:negativeparity_eigenphase}. The ANCs are presented for a given spin-parity state of $ {}^{9}\mathrm{Be} $, relative orbital momentum of the neutron with respect to the $ {}^{9}\mathrm{Be} $ cluster, and total spin of the $ \mathrm{n} + {}^{9}\mathrm{Be} $ partition. The only column shown contains the ANCs obtained by phenomenological adjustment of the NCSMC eigenenergies to their experimental values, given that the $ 2^{-} $ is near-threshold but unbound in the raw calculation.}
\end{table}

Before proceeding to the positive parity states, we present the ANCs -- see Eq.~(\ref{eq:ANC}) for their definition -- for the $ 1^{-} $ and $ 2^{-} $ halo states of $ {}^{10}\mathrm{Be} $ in Tables~\ref{section:10Be:table:ANCs_for_1m} and~\ref{section:10Be:table:ANCs_for_2m}, respectively, using the same interaction and configuration space as in determination of the phase shifts in Fig.~\ref{section:10Be:figure:negativeparity_eigenphase}. In Table~\ref{section:10Be:table:ANCs_for_1m}, we provide the ANCs for the $ 1^{-} $ state of $ {}^{10}\mathrm{Be} $ for each partial wave channel of the $ \mathrm{n} + {}^{9}\mathrm{Be} $ partition. These are labeled by the spin-parity of the relevant $ {}^{9}\mathrm{Be} $ state, the relative orbital momentum of the neutron with respect to the $ {}^{9}\mathrm{Be} $ cluster, and the total spin of the coupled $ \mathrm{n} + {}^{9}\mathrm{Be} $ partition. In the first column with ANC results, we show the raw output of the NCSMC from the aforementioned calculations. Due to the dependence of the asymptotic wave function on $ \kappa \propto \sqrt{-E} $, the ANCs are incredibly sensitive to the prediction for the energy eigenvalues. Thus, by phenomenologically adjusting the NCSMC energy eigenvalues to match the experimental ones, one can get a more realistic picture of the ANCs. That is what is then shown in the second ANC column, labeled as \textit{ANC Pheno}. One can obviously see the dramatic difference in values for the predicted ANCs coming from relatively minor shifts in the overall energy eigenvalues, typically of order $ 1 \  \mathrm{MeV} $ or so. Identically, in Table~\ref{section:10Be:table:ANCs_for_2m} we show the ANCs for the $ 2^{-} $ state of $ {}^{10}\mathrm{Be} $. In this case, there are no ANCs for the raw NCSMC calculation since the state is found to be an unbound, near-threshold resonance, so we show only the \textit{ANC Pheno} results coming from phenomenological adjustment of the $ 2^{-} $ state. Based upon the extracted values for the ANCs in both the $ 1^{-} $ and $ 2^{-} $ states, we find dominance of the $ l = 0 $ channels which further supports the interpretation of these states as excited halo candidates, regardless of any additional sub-structure of the $ {}^{9}\mathrm{Be} $ partition. While in both states the dominant structure arises from the $ {}^{9}\mathrm{Be}[ 3/2^{-} ] \, \otimes \, n $ channel, there are still noteworthy contributions from the $ {}^{9}\mathrm{Be}[ 5/2^{-} ] \, \otimes \, n $ in both states, and $ {}^{9}\mathrm{Be}[ 1/2^{-} ] \, \otimes \, n $ in the $ 1^{-} $ state. Then, the resulting picture is therefore not that of a single-channel halo state with no internal clustering, but rather of a multi-channel system whose asymptotics are nevertheless dominated by $ \mathrm{S} $-wave neutron motion around a $ {}^{9}\mathrm{Be} $ core.

In Fig.~\ref{section:10Be:figure:positiveparity_eigenphase}, we find a wealth of positive parity resonance states compared to the case of negative parity, as is characteristic of our theoretical~\cite{BJP.49.5766.2022, PRC.66.044310.2002} and experimental understanding of the rather dense $ {}^{10}\mathrm{Be} $ spectrum. Importantly, we find that the three lowest-lying bound states as predicted by the NCSMC, that is, the $ 0^{+}_{1} $ sitting at $ -5.9721 \ \mathrm{MeV} $, the $ 2^{+}_{1} $ at $ -2.5104 \ \mathrm{MeV} $, and the $ 2^{+}_{2} $ at $ -0.6646 \ \mathrm{MeV} $, are notably all bound to the correct experimental values within an $ \mathrm{MeV} $ or so. For reference, these are respectively $ -6.8122 \ \mathrm{MeV} $, $ -3.444 \ \mathrm{MeV} $, and $ -0.8538 \ \mathrm{MeV} $~\cite{TUNL.Be10.2004}. The corresponding partial wave channels are the $ {}^{3}\mathrm{P}_{0} $, $ {}^{3}\mathrm{P}_{2} $, and $ {}^{5}\mathrm{P}_{2} $. As mentioned prior, we do not bind the excited state $ 0^{+}_{2} $ due to lacking alpha clustering effects, though we make note of a $0^+$ state appearance as a proper resonance of the $ \mathrm{n} + {}^{9}\mathrm{Be} $ system via the $ {}^{3}\mathrm{P}_{0} $ channel seen on the right-hand-side of Fig.~\ref{section:10Be:figure:positiveparity_eigenphase}. Beyond inclusion of the additional $ {}^{6}\mathrm{He} \, + \, \alpha $ partition, as discussed in Ref.~\cite{NPA.958.78100.2017}, further effects of cluster polarization likely play a significant role in the formation of bound and resonant states in $ {}^{10}\mathrm{Be} $.

On to the scattering states, the NCSMC predicts the existence of a near-threshold $ 1^{+} $ resonance, already seen in earlier NCSM calculations~\cite{Caurier2002}, which has yet to be confirmed experimentally. This is consistent with calculations of $ {}^{10}\mathrm{C} $ and $ {}^{10}\mathrm{B} $~\cite{gennari2021ab}, and is further anticipated to exist based on the observed isospin analogue states in the spectra of those nuclei. A second narrow and unobserved $ 1^{+} $ resonance is also predicted about $ 3 \ \mathrm{MeV} $ higher in energy. While most of the eigenphase shifts correspond to a single dominant partial wave channel, the $ 1^{+}_{2} $ is an exception and comes from stimulation of the same partial wave channel as the $ 1^{+}_{1} $ at higher c.m. energy, as can be seen from the behavior of the $ {}^{5}\mathrm{P}_{1} $ phase shift. Moving up in spin, we find a $ 2^{+}_{3} $ resonance built from the $ {}^{5}\mathrm{P}_{2} $ partial wave which we cannot readily match to an experimentally observed resonance state. There exists a resonance with energy $ 0.7298 \ \mathrm{MeV} $ which is presumably related, though our calculation is about $ 2 \ \mathrm{MeV} $ above this and thus we cannot say for certain. We observe two additional $ 2^{+} $ resonances at higher c.m. energy coming from the $ {}^{7}\mathrm{P}_{2} $ and $ {}^{3}\mathrm{P}_{2} $ channels, though they are not readily discerned as any particular state seen in the known experimental spectrum. Lastly, a very narrow $ 3^{+} $ resonance is observed at quite low c.m. energy in the $ {}^{5}\mathrm{P}_{3} $ channel, a place which -- when referring to the experimental spectra~\cite{TUNL.Be10.2004} -- is seemingly empty of resonance states.

\subsection{Borromean halo nucleus $^6$He}\label{subsec:He6}
A stringent test of any \emph{ab initio} description of three-cluster dynamics is provided by Borromean halo
nuclei, where the bound ground state emerges only through genuine three-body correlations and the wave
function exhibits pronounced long-range asymptotics. The $^6$He nucleus  is a prototypical example,
with a weakly bound ground state and an extended spatial distribution that reflects its dominant
$\alpha+n+n$ character. As such, it offers an ideal benchmark to assess whether a unified treatment can
simultaneously describe both short-range many-body correlations and the correct three-body continuum
behavior.

An initial \emph{ab initio} description of ${}^{6}\mathrm{He}$ and its $\alpha+n+n$ continuum was obtained 
in a model space spanned only by microscopic cluster channels, i.e.\
by omitting discrete NCSM eigenstates of the composite system in the wave function ansatz~\cite{Quaglioni2013,Romero2014}.
This three-cluster treatment captured the correct three-body asymptotics and enabled continuum calculations;
however, it was clear that additional short-range many-body correlations were missing and that convergence with
respect to the model space was comparatively slow. This work was followed by a full NCSMC description, 
also including square-integrable NCSM eigenstates of the ${}^{6}\mathrm{He}$ system
and thus accelerating convergence for bound-state and low-energy continuum observables~\cite{Romero2016,Quaglioni2018}.
The three-cluster NCSMC formalism was demonstrated and quantified through a detailed study of the ${}^{6}\mathrm{He}$
ground state and low-lying continuum, using SRG-evolved chiral N$^{3}$LO NN interactions at two resolution scales,
$\lambda_{\mathrm{SRG}}=1.5~\mathrm{fm}^{-1}$ and $2.0~\mathrm{fm}^{-1}$, while omitting initial and induced 3N forces
in those calculations~\cite{Quaglioni2018,Romero2016}. 

\begin{figure}[H]
\includegraphics[width=8cm]{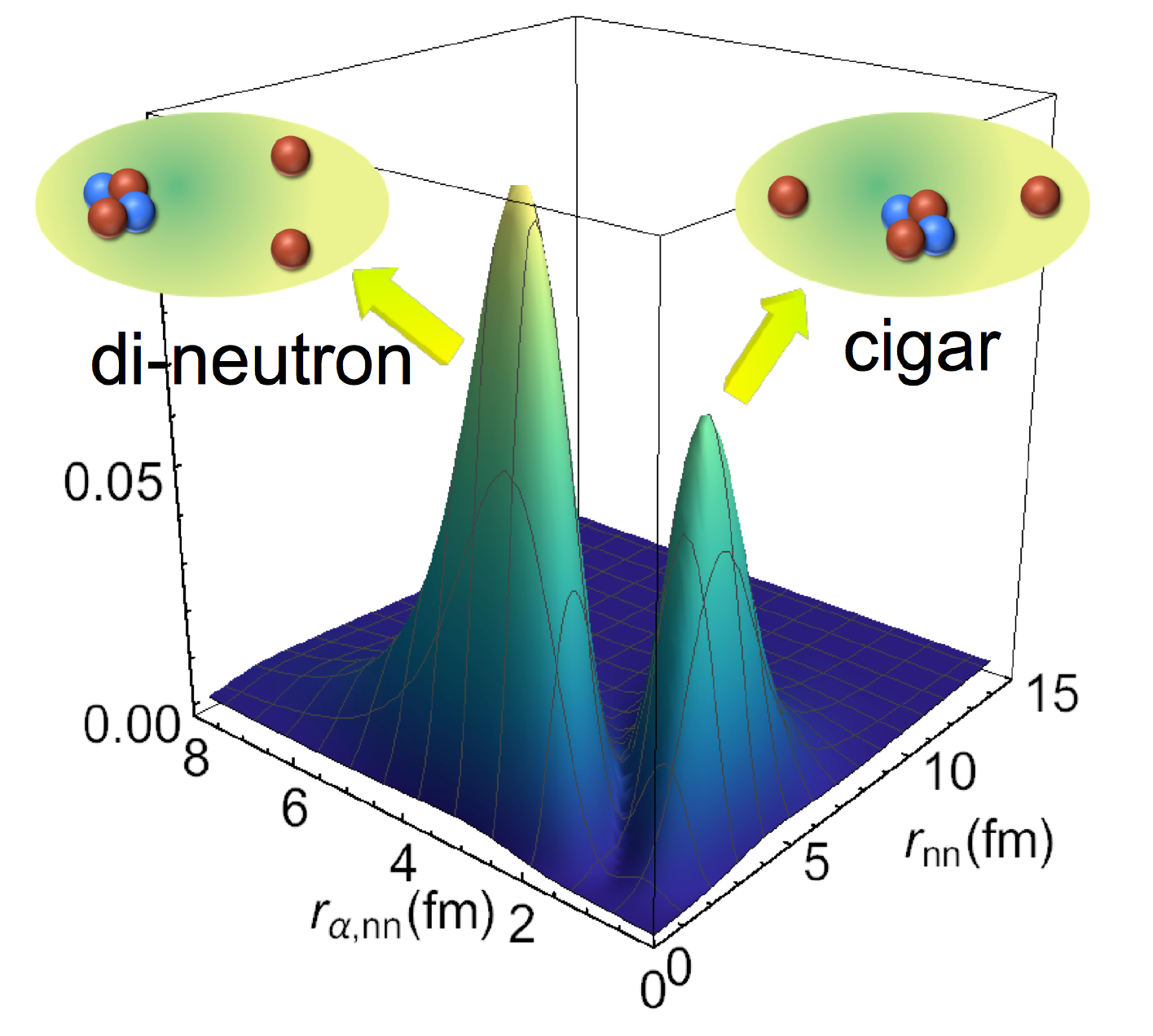}
\caption{Probability distribution the $J^\pi = 0^+$ ground state of the $^6$He. Here $r_{nn} = \sqrt{2} \eta_{nn}$ and $r_{\alpha,nn} = \sqrt{3/4} \eta_{\alpha,nn}$ are, respectively, the distance between the two neutrons and the distance between the c.m. of $^4$He and that of the two neutrons.}
\label{fig:He6_gs-structure}
\end{figure}
\begin{figure}[t]
\centering
\includegraphics[width=16cm]{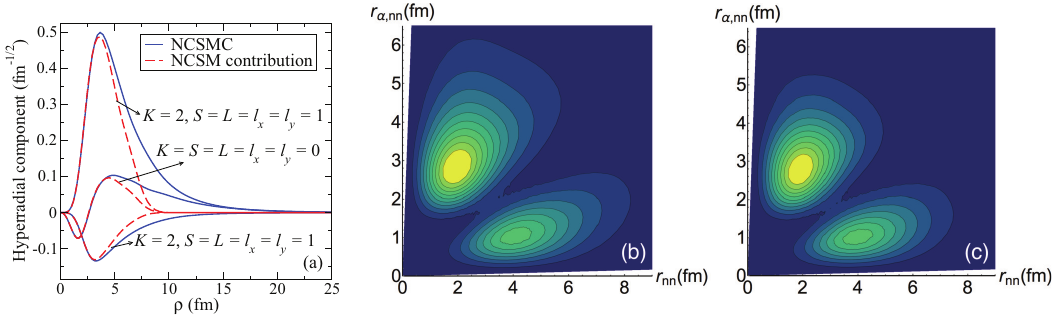}
\caption{(a) Most relevant hyper-radial components of the $\alpha + n + n$ relative motion within the $^6$He g.s. after projection of the full NCSMC wave function (blue solid lines) as well as of its NCSM portion (red dashed lines) into the orthogonalized microscopic-cluster basis. [(b), (c)] Contour plots of the probability distribution obtained from the projection of the full NCSMC wave function of panel (a) and its NCSM component, respectively, as a function of the relative coordinates $r_{nn} = \sqrt{2} \eta_{nn}$ and $r_{\alpha,nn} = \sqrt{3/4} \eta_{\alpha,nn}$. Figure from Ref.~\cite{Quaglioni2018}.}
\label{fig:He6_cluster_fromf}
\end{figure}

For the $\lambda_{\mathrm{SRG}}=2.0~\mathrm{fm}^{-1}$ interaction, the NCSMC calculation yields a realistic
${}^{6}\mathrm{He}$ ground-state energy of -29.17 MeV (compared to the experimental value of -29.268 MeV~\cite{Brodeur:2012zz}). 
The corresponding g.s.\ wave function exhibits the expected dineutron-dominated spatial distribution when analyzed through the
three-body probability density constructed from projecting the full NCSMC
solution onto the microscopic three-cluster basis (Fig.~\ref{fig:He6_gs-structure}).  
Beyond this qualitative picture, the same continuum-coupled description yields
converged matter ($r_m$) and point-proton radii ($r_{pp}$) in computationally accessible model spaces. 
More importantly, it enables simultaneously a description of the small two-neutron separation energy ($S_{2n}=0.94(5)$ MeV) and the extended spatial size of ${}^{6}\mathrm{He}$ ($r_m = 2.46(2)$ fm, $r_{pp}=1.90(2)$ fm) broadly consistent with experimental constraints. 
This is notable, given that in traditional \textit{ab initio} calculations limited to expansions on square-integrable basis states, including the NCSM, the matter and point-proton radii of ${}^{6}\mathrm{He}$ converge slowly with model-space increase, reflecting the difficulty of representing the long-range halo tail. This difference is exemplified by Fig.~\ref{fig:He6_cluster_fromf}, which compares the hyper-radial components of the
$\alpha+n+n$ relative motion in the ${}^{6}\mathrm{He}$ ground state after projection of the full NCSMC wave function and of its NCSM portion onto the orthogonalized microscopic-cluster basis. The
comparison makes apparent the deficiency of the square-integrable NCSM component in reproducing the long-range halo tail,
and how the inclusion of explicit three-cluster continuum degrees of freedom in the NCSMC restores the correct extended
behavior.

\begin{figure}[H]
\centering
\includegraphics[width=9cm]{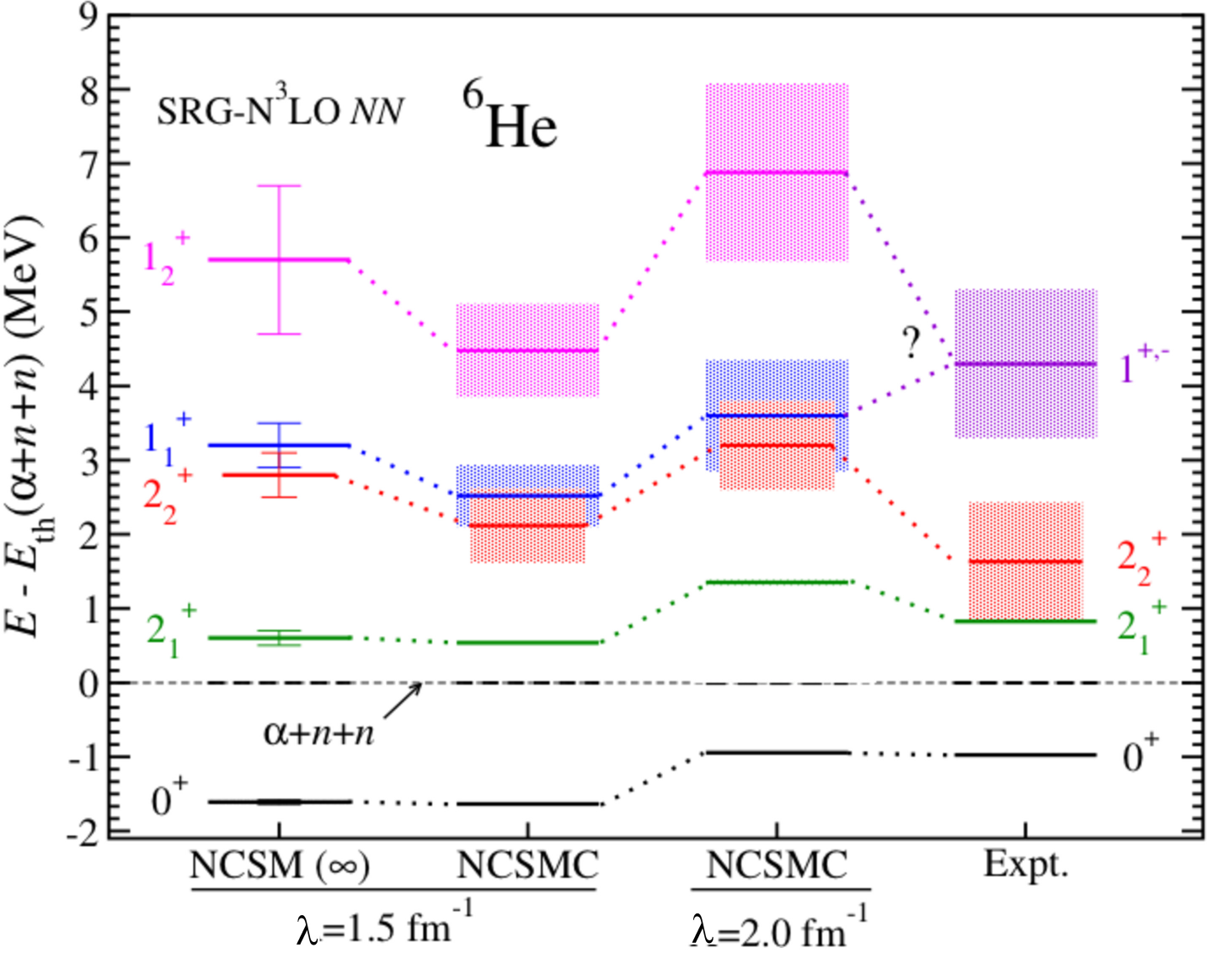}
\caption{Spectrum of low-lying energy levels of the $^6$He nucleus. Results for
the SRG-N$^3$LO NN interaction with $\lambda=1.5$ fm$^{-1}$ are shown on the left-hand side for both NCSM and NCSMC. 
The third set of energy levels corresponds to NCSMC results obtained with $\lambda=2.0$ fm$^{-1}$  and 
the fourth to the experimental spectrum of Ref.~\cite{Mougeot:2012aq}. Figure from Ref.~\cite{Romero2016}.}
\label{fig:He6_spectrum}
\end{figure}

As a further illustration of the role of continuum degrees of freedom and many-body correlations, Fig.~\ref{fig:He6_spectrum}
shows the low-lying ${}^{6}\mathrm{He}$ spectrum obtained with the SRG-evolved N$^{3}$LO NN interaction at
$\lambda=1.5$ and $2.0~\mathrm{fm}^{-1}$. There, we compare the NCSMC results of Ref.~\cite{Quaglioni2018} with the NCSM spectrum obtained by treating the ${}^{6}\mathrm{He}$ excited states as bound states.
Besides the results in the largest accessible HO model space ($N_{\max}=12$), for the NCSM we also show the spectrum
extrapolated to the infinite-space limit.
Because the NCSM is a bound-state technique and does not yield resonance widths, only the excitation energies (with the
estimated extrapolation uncertainty) are shown in that case, whereas for the NCSMC the resonances are represented
by their centroids (solid lines) and widths (shaded areas).
While very narrow resonances such as the first $2^{+}$ can be captured reasonably within the
bound-state approximation, the description of broader states generally requires both short-range many-body
correlations and explicit coupling to the three-body continuum.

The two SRG resolution scales yield a qualitatively similar pattern, and their differences provide a rough estimate of
the impact of omitted induced three-nucleon (and higher-body) interactions, which are needed to restore the formal
unitarity of the SRG transformation. More generally, explicit 3N forces (including the initial chiral 3N interaction) are indispensable for an accurate description of the spectrum as a whole. Indeed, while the SRG-evolved NN interaction at $\lambda=2.0~\mathrm{fm}^{-1}$ yields a realistic energy and structure for the ${}^{6}\mathrm{He}$ ground state, neither of the two adopted resolution scales reproduces quantitatively the
low-energy excited spectrum reported in Ref.~\cite{Mougeot:2012aq}.

\subsection{Large-scale NCSM calculations for $^{11}$Li}\label{subsec:Li11}

$^{11}$Li is the nucleus where a neutron halo was discovered in interaction cross section measurements more than four decades ago~\cite{TA85}. This drip-line isotope has two weakly bound neutrons in its ground state with separation energy $S_{2n}$ = 369.2(6) keV that have a large spatial extent compared to the core nucleus $^9$Li leading to a $^{9}$Li + $n$ + $n$ three-body Borromean halo. This exotic structure was postulated to give rise to unconventional excitation modes. It was predicted that a low-energy dipole resonance could arise due to oscillation of the weakly bound halo neutrons against the core \cite{Ikeda1992}. A non-resonant soft electric dipole mode in a Coulomb excitation process was also predicted \cite{HansenJonson1987}. There have been several experiments performed aimed at elucidating the nature of its ground state as well as its excitation modes~\cite{Tanihata1996,TANIHATA2013215,Bohlen1995,Korsheninnikov1996,Korsheninnikov1997,Gornov1998,Simon2007,Tanihata2008,Kanungo2015,Korotkova2015,Tanaka2017}. Simultaneously, understanding properties of $^{11}$Li have been subject to numerous theoretical studies~\cite{Barranco2001,Ershov2004,Hagino2005,Hagino2007,Redondo2008,Hagino2009,Potel2010,Kikuchi2013}. Yet, an \textit{ab initio} description of this complex system is still lacking.

As a step in the direction of remedying this situation, we have performed \textit{ab initio} calculations of $^{11}$Li nuclear structure using the NCSM approach. As input, we employed the chiral EFT NN and 3N interaction NN-N$^4$LO + 3N$^*_\mathrm{lnl}$ (denoted as NN N$^4$LO + 3N$_\mathrm{lnlE7}$ in the figures). The interaction has been softened by the SRG technique with the SRG induced three-nucleon terms fully included. The evolution parameter $\lambda_{\rm SRG}=1.8$ fm$^{-1}$ has been used primarily and we have checked that observables are insensitive to the variation of the $\lambda_{\rm SRG}$ parameter between 1.8 and 2.0 fm$^{-1}$. For earlier NCSM studies reporting some $^{11}$Li results obtained using NN interactions only, see Refs.~\cite{Navratil1998,Forssen2009,Caprio2022,johnson2025challengesfirstprinciplesnuclearstructure}. 

%
\begin{figure}[H]
\centering
\subfloat[\centering]{\includegraphics[width=8.4cm]{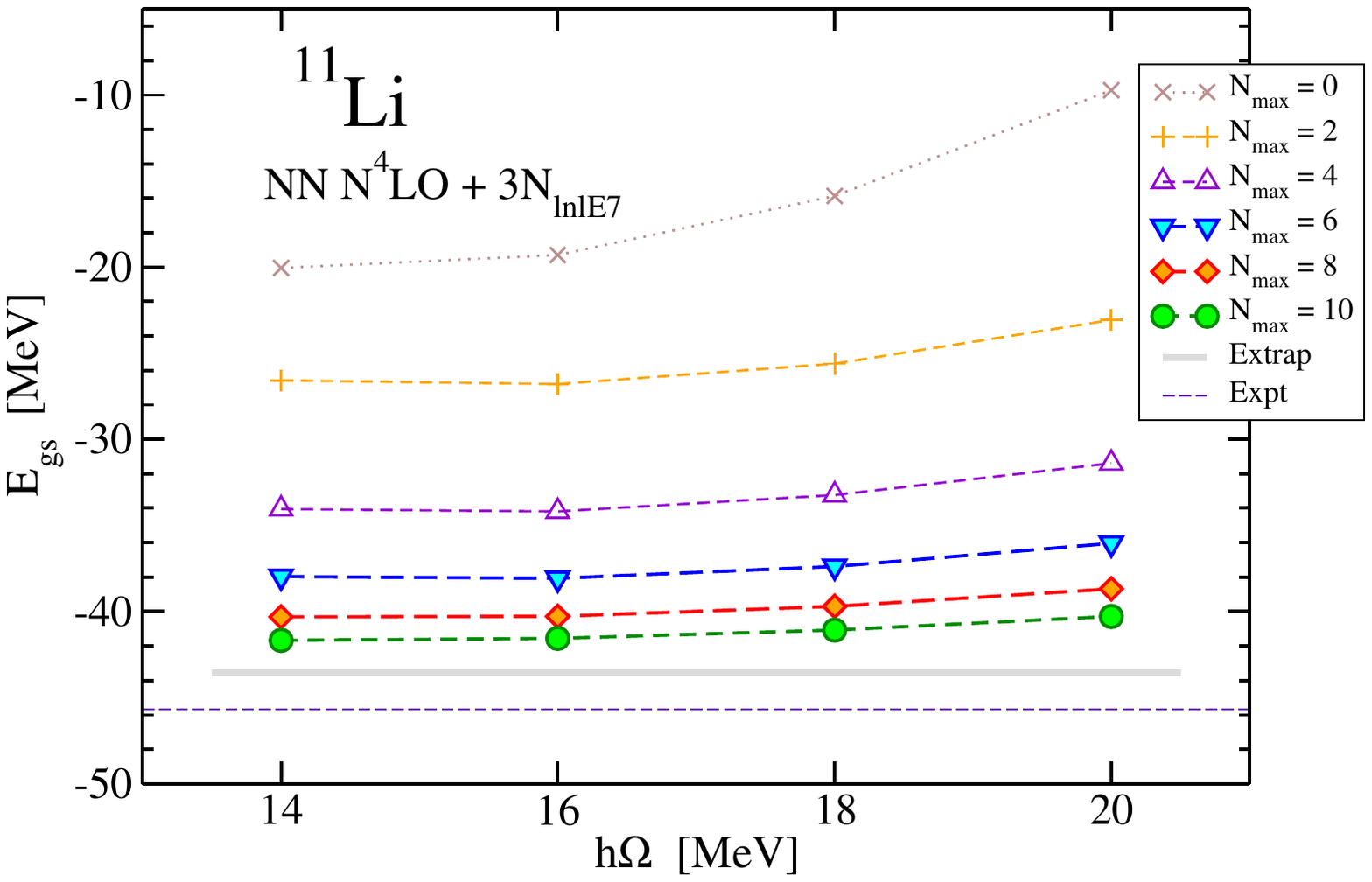}}
\subfloat[\centering]{\includegraphics[width=8.4cm]{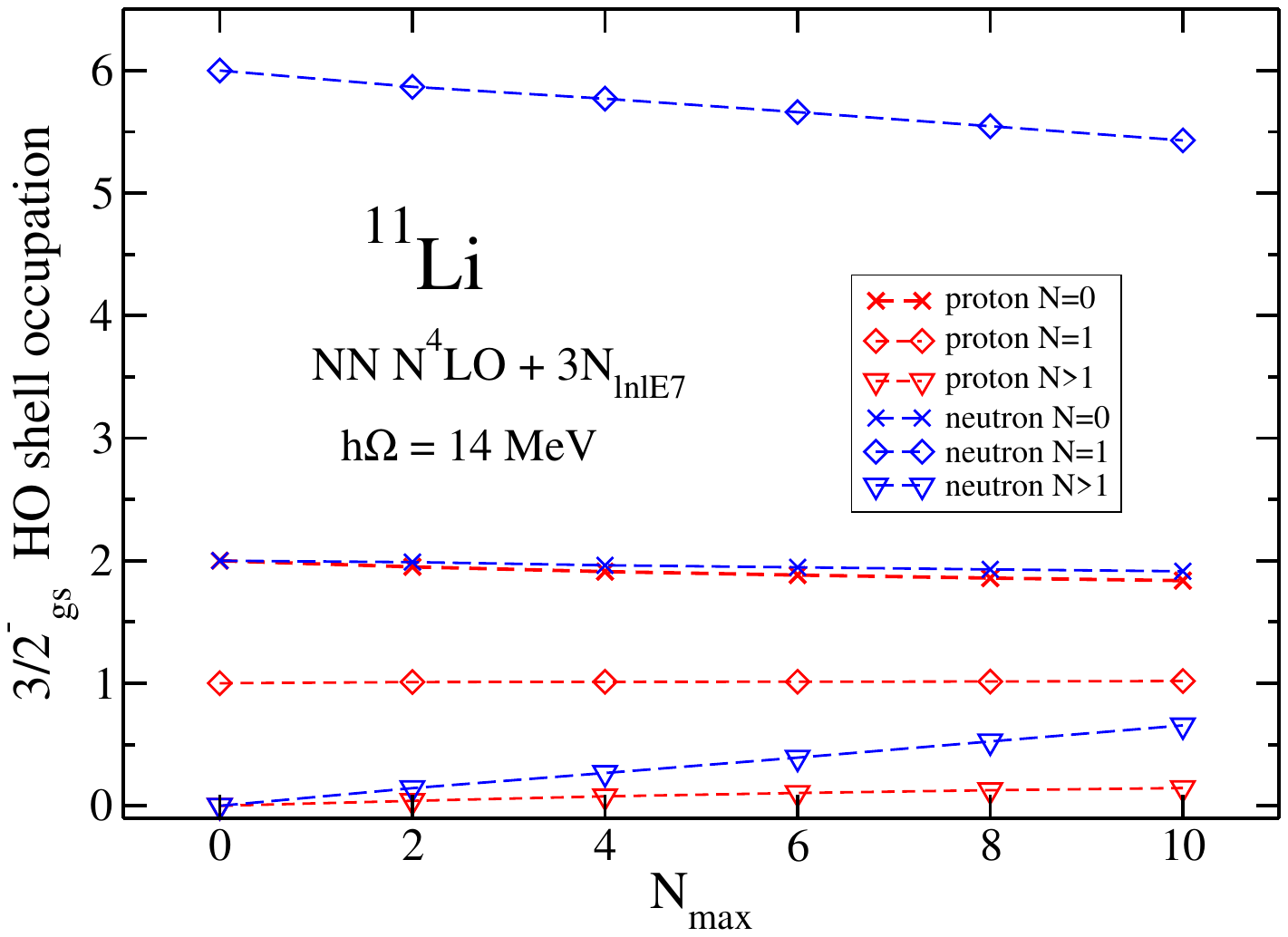}}
\caption{(a) $^{11}$Li $3/2^-$ ground-state energy dependence on the harmonic oscillator frequency for different NCSM model space sizes characterized by
$N_{\rm max}$. The gray band shows the $N_{\rm max}\rightarrow\infty$ extrapolated value with its uncertainty. The dotted lines correspond to the experimental ground-state energies. (b) The ground-state HO shell occupation dependence on $N_{\rm max}$ for the $\hbar\Omega{=}14$ MeV calculation. The SRG-evolved NN-N$^4$LO + 3N$^{*}_{\rm lnl}$ interaction was used with the HO frequency of $\hbar\Omega{=}14$ MeV. See the text for further details.}
\label{fig:Li11_Egs_occprob}
\end{figure}
In panel (a) of Fig.~\ref{fig:Li11_Egs_occprob}, we present the $^{11}$Li $3/2^-$ ground-state energy dependence on the NCSM HO frequency in the range of $\hbar\Omega{=}14{-}20$ MeV for the basis size up to $N_{\rm max}{=}10$. The basis dimension reaches 929 million at $N_{\rm max}{=}10$.  The extrapolated results to $N_{\rm max}{\rightarrow}\infty$ using the exponential function $E(N_{\rm max})=a + b\,e^{-cN_{\rm max}}$ are shown by the gray band. The uncertainties are obtained by varying the number of extrapolated points, the HO frequencies and the SRG evolution parameter. In addition, we have performed the same for $^9$Li, see Ref.~\cite{Singh2026}. The predicted ground-state energy of $^{11}$Li is $-43.56 (35)$ MeV while that of $^9$Li is $-43.73 (18)$ MeV that can be compared to experimental $-45.709$ MeV and $-45.34$ MeV, respectively. Overall, we find a slight underbinding of $\sim 1.5 - 2$ MeV and within uncertainties about the same ground-state energy of the two isotopes. We find that the experimentally well-bound $^9$Li ground-state energy converges faster while the very weakly bound $^{11}$Li would benefit from an inclusion of three-body cluster components in the trial wave function absent in the present NCSM calculations that might help to bind it with respect to $^9$Li.

In panel (b) of Fig.~\ref{fig:Li11_Egs_occprob}, we show the occupations of the major HO shells for the $^{11}$Li $3/2^-$ ground-state as they evolve with the basis size enlargement. While the proton occupations remain stable, the neutron occupation of the $N{=}1$ (0$p$-shell) decreases and the neutron occupation of the higher $N$ shells steadily increases with $N_{\rm max}$.  

%
\begin{figure}[H]
\centering
\subfloat[\centering]{\includegraphics[width=8.2cm]{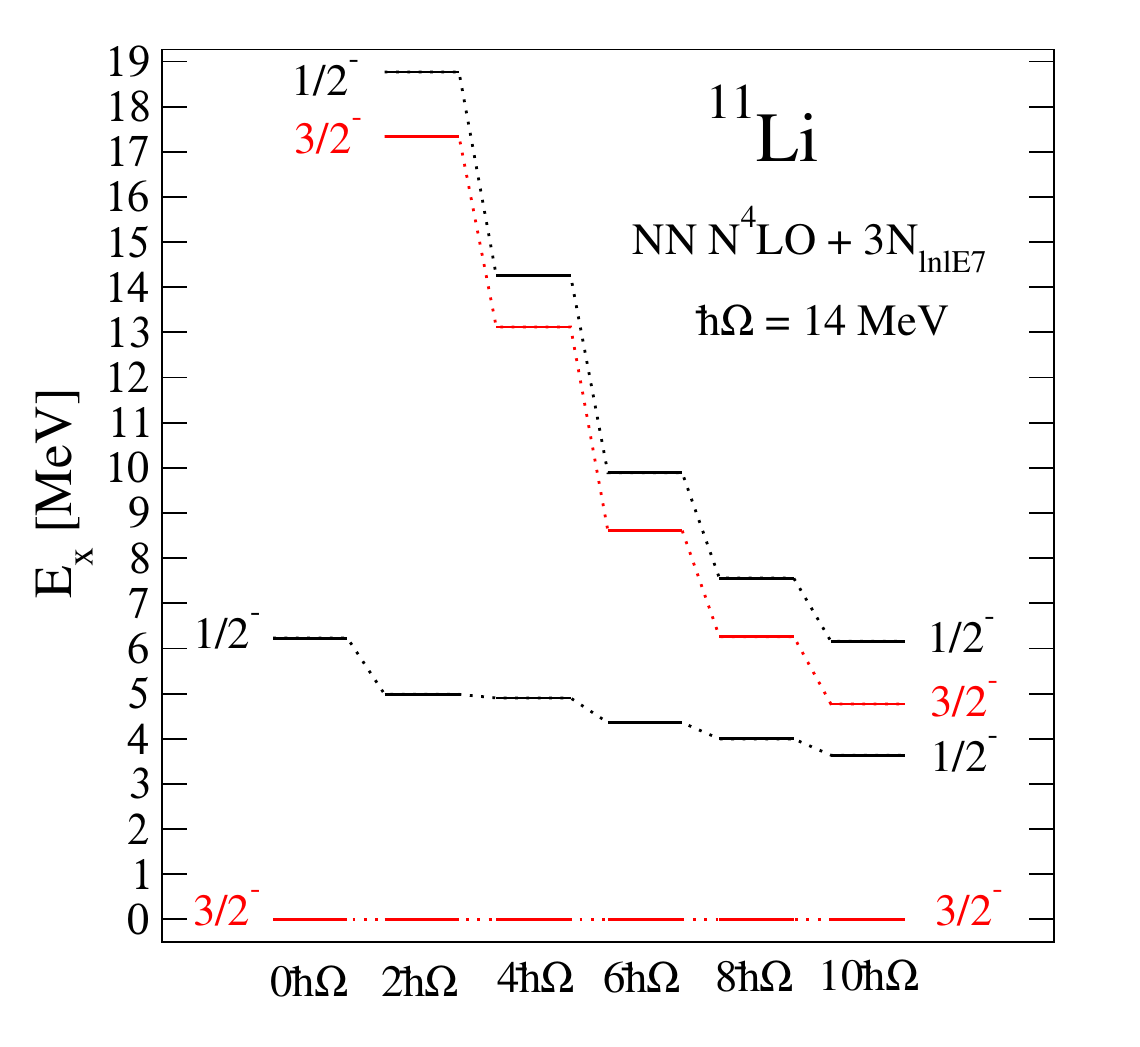}}
\subfloat[\centering]{\includegraphics[width=8cm]{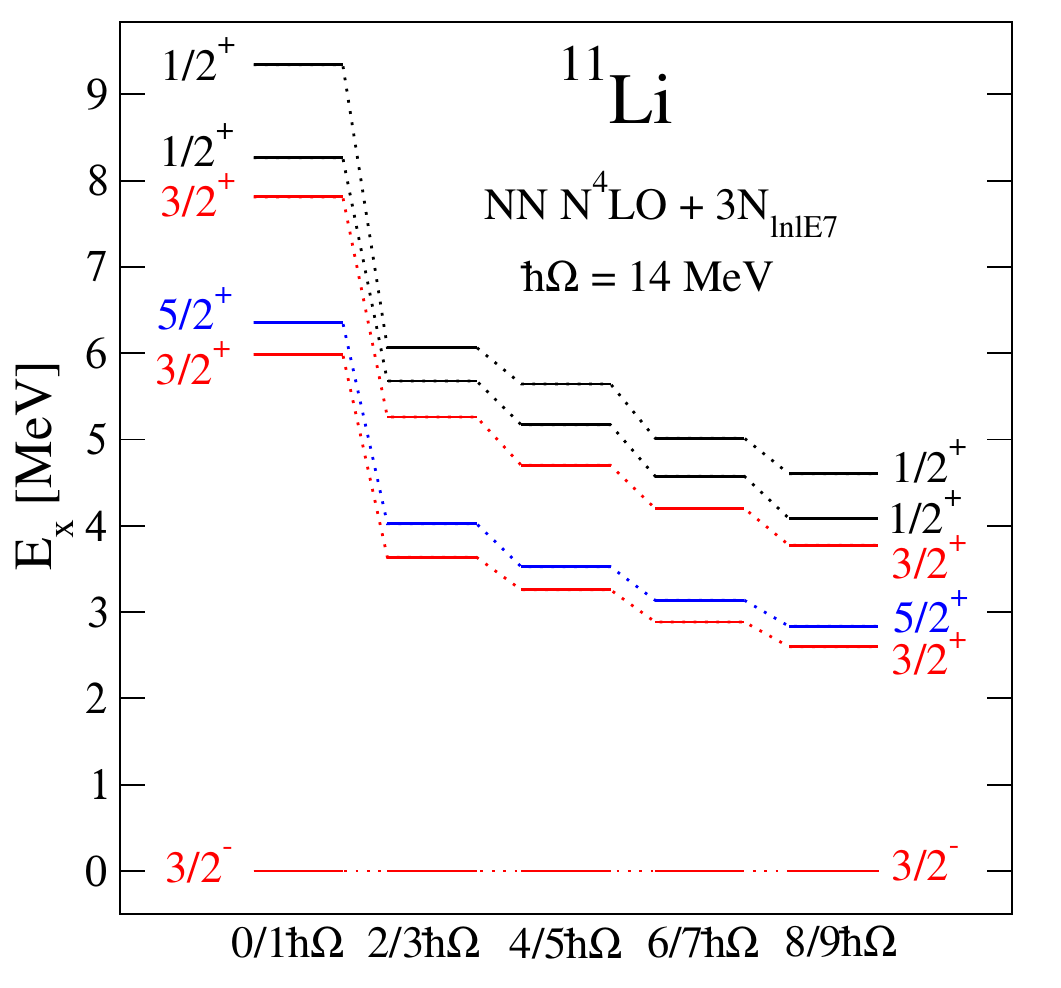}}
\caption{Excitation energy dependence on the NCSM basis size for the low-lying negative-parity (panel (a)) and positive-parity (panel (b)) states of $^{11}$Li. The SRG-evolved NN-N$^4$LO + 3N$^{*}_{\rm lnl}$ interaction was used with the HO frequency of $\hbar\Omega{=}14$ MeV. See the text for further details.}
\label{fig:Li11_Eexc}
\end{figure}
The dependence of the lowest calculated excited states on the NCSM basis size are shown in Fig.~\ref{fig:Li11_Eexc} for negative-parity states, panel (a), and for the positive-parity states, panel (b). In the latter, the negative-parity $3/2^-$ ground state obtained in an $N_{\rm max}$ basis space is matched with the positive-parity excited states obtained in the $N_{\rm max}{+}1$ basis space. For technical reasons, the largest positive-parity space we are able to reach is $N_{\rm max}{=}9$ with the dimension of 269 million. While the $1/2^-_1$ state excitation energy decreases gradually with $N_{\rm max}$, the multi-$\hbar\Omega$ dominated $3/2^-_2$ and $1/2^-_2$ states manifest a rapid decrease of excitation energies with the basis size correlated with an increase of higher-$\hbar\Omega$ components in the $3/2^-_1$ ground-state wave function, signature of which is seen in Fig.~\ref{fig:Li11_Egs_occprob} (b). This trend was also observed in recent large-scale NCSM calculations using NN interactions only~\cite{johnson2025challengesfirstprinciplesnuclearstructure}. Similarly, the positive-parity state excitation energies decrease steadily with $N_{\rm max}$. As seen in Fig~\ref{fig:Li11_Eexc}, we find the lowest positive-parity states, $3/2^+_1$ and $5/2^+_1$, below the lowest negative-parity excited states in the largest spaces we could reach. 

It should be noted that in experiment, only the $^{11}$Li $3/2^-_1$ ground state is bound. The excited states calculated within the NCSM could approximate resonances in the continuum. To establish connection to experimental observations, we have investigated multipole operator transitions from the ground state to excited states with details given in Ref.~\cite{Singh2026}.

The present large-scale NCSM calculations are a prerequisite of planned investigation of $^{11}$Li within the NCSMC treating this halo nucleus as a three-body cluster system of $^9$Li and two neutrons that will be capable providing a realistic description  of the two-neutron halo similarly as accomplished for $^6$He~\cite{Quaglioni2018}, see Subsection~\ref{subsec:He6}.

\section{Discussion}\label{sect:discuss}

In this article, we provided evidence of usefulness and power of \textit{ab initio} nuclear theory for the description and understanding of halo nuclei, exotic weakly bound systems with extended single-nucleon or two-neutron density beyond a tightly bound core. We reviewed applications of the no-core shell model with continuum, a method that provides a unified description of bound and unbound nuclear states starting from precision chiral EFT based NN+3N interactions, to single-neutron halo nuclei $^{11}$Be and $^{15}$C, a single-proton halo nucleus $^8$B, as well as to $^6$He exhibiting two-neutron Borromean halo. Further, we provided an analysis of excited halo states in $^{10}$Be. 

The NCSMC is currently the only method that has successfully reproduced the parity inversion in the ground state of $^{11}$Be and, at the same time, obtained wave functions of its $1/2^+$ and $1/2^-$ bound states demonstrating the neutron+$^{10}$Be($0^+$) $S$ wave and $P$ wave halos, respectively, that extend well beyond 20 fm. Calculated ANCs of the two halo states agree very well with those extracted from the knockout and transfer reaction halo-EFT analyses~\cite{Yang2018,Hebborn2021}.

We presented new NCSMC results for the single-neutron halo nucleus $^{15}$C. Focusing on the description of its two bound states, the $^{14}$C+n $S$ wave halo $1/2^+$ ground state and the weakly bound $5/2^+$ excited state, we showed a good agreement of the calculated ANCs with those obtained by analysis of $^{14}$C(d,p)$^{15}$C transfer and knockout reactions~\cite{Mukhamedzhanov2011,Moschini2019,Hebborn2021}. Also, we performed calculations of the $^{14}$C(n,$\gamma$)$^{15}$C capture reaction cross section relevant for several astrophysical processes. Our results, obtained for both bound final states, are in line with recent experimental measurements~\cite{Reifarth2008} and phenomenological calculations~\cite{Tkachenko2025}, although slightly higher than the most recent experimental determination~\cite{Jiang_2025}.

We discussed the structure of the very weakly bound single-proton halo nucleus $^8$B that plays an important role in astrophysics. We provided a detailed analysis of its $2^+$ halo ground state dominated by the $^7$Be($3/2^-$)+proton cluster in the relative $P$ wave. The ANCs obtained within the NCSMC are in good agreement with the recent experimental determination~\cite{Paneru2019}.

While the $^{10}$Be nucleus is well bound, it features two excited states of halo nature just below the $^9$Be+neutron threshold. In new calculations, we investigated the $^{10}$Be bound and scattering states within NCSMC focusing particularly on the structure of the two excited halo states, $1^-$ and $2^-$, dominated by $^9$Be($3/2^-$)+neutron in the $S$ wave with spins anti-aligned and aligned, respectively. We provided a detailed analysis of the two states with ANCs for various partial waves. In addition, we discussed resonances above the $^9$Be+neutron threshold where we predict, e.g., a $1^+$ resonance not included in the recent data evaluation~\cite{TUNL.Be10.2004}. The present calculations will be improved in the future by coupling the $^6$He+$\alpha$ mass partition that lies experimentally just 600 keV above the $^9$Be+neutron threshold, thus impacting the structure of the low-lying positive-parity resonances.

The NCSMC has been extended recently to describe systems dominated by three-body breakup channels. Applying this formalism, we discussed the structure of the two-neutron Borromean halo nucleus $^6$He described within the NCSMC as an $\alpha$+n+n system. The weakly bound $0^+$ ground states manifests a superposition of the di-neutron and cigar configurations as found in earlier cluster model calculations, here obtained microscopically from realistic nucleon-nucleon interactions. The NCSMC calculations also predict several resonances above the $\alpha$+n+n threshold starting with a narrow $2^+$ state corresponding to the experimentally well established first exited state of $^6$He. This is then followed by broad $2^+$ and $1^+$ resonances with the former one could be matched to experimentally observed broad $2^+$ state~\cite{Mougeot:2012aq}. Due to the complexity of the three-body cluster NCSMC calculations, the $^6$He results have so far been obtained using the NN interaction only, with the 3N interaction capability still to be implemented.

The next challenge for the NCSMC and for the \textit{ab initio} nuclear theory in general, is the description of the two-neutron Borromean halo nucleus $^{11}$Li, the exotic system discovered four decades ago~\cite{TA85} later interpreted as the first ever halo nucleus found~\cite{HansenJonson1987}. It poses a significant challenge and added complexity compared to the achieved NCSMC $^6$He calculations due to its heavier mass, non-zero spin, and the necessity to include excited state(s) of the $^9$Li core. As the required first step, we have presented here large-scale no-core shell model calculations of $^{11}$Li reaching basis spaces up to $N_{\rm max}{=}10$ that allow extrapolating the total binding energy. We have calculated the same for $^9$Li and found that NCSM predicts both nuclei bound by about the same energy close to experimental one. It is reasonable to anticipate that by including the $^9$Li+n+n continuum within the three-body cluster NCSMC, the $^{11}$Li would become bound. We have also discussed excited states $^{11}$Li predicted by NCSM. In the largest spaces reached, we find the lowest positive-parity states below the lowest negative-parity excited states, see Ref.~\cite{Singh2026} for further details. It will be also very important to investigate the excited states within the NCSMC, i.e., including $^9$Li+n+n continuum. 


\vspace{6pt} 




\funding{This work was supported by the Natural Sciences and Engineering Council of Canada (NSERC) Grant No. SAPIN-2022-00019. Part of this work was performed under the auspices of the U.S. Department of Energy by Lawrence Livermore National Laboratory under contract DE-AC52-07NA27344 and under Contract DE-AC52-07NA27344. TRIUMF receives federal funding via a contribution agreement with the National Research Council of Canada. Computing support came from an INCITE Award on the Frontier supercomputer of the Oak Ridge Leadership Computing Facility (OLCF) at ORNL, from the Lawrence Livermore National Laboratory (LLNL) Institutional Computing Grand Challenge program, and from the Digital Research Alliance of Canada.}


\acknowledgments{We acknowledge the organizers of the conference “International Symposium Commemorating the 40th Anniversary of the Halo Nuclei (HALO-40)” for the contribution invite.}





\isPreprints{}{
} 

\reftitle{References}



\isPreprints{}{
} 
\end{document}